\documentclass[12pt]{article}
\usepackage{graphicx}
\usepackage{multirow}

\usepackage{algorithm}
\usepackage{algpseudocode}

\usepackage{amsmath,amssymb,amsfonts,nicefrac,mathtools,bbm,amsthm}
\newcommand{\BIT}{\begin{itemize}}
\newcommand{\EIT}{\end{itemize}}
\newcommand{\BNUM}{\begin{enumerate}}
\newcommand{\ENUM}{\end{enumerate}}
 
\newcommand\mbb[1]{\mathbb{#1}}

\def\reals{\mathbb{R}} 
\def\indic#1{\mbb{I}\left({#1}\right)} 

\newtheorem{lemma}{Lemma}

\newcommand{\putFig}[3]{
	\begin{figure}[h!]
		\centering
		\includegraphics[width=#3]{figs/#1}
		\caption{#2}
		\label{fig:#1}
	\end{figure}
}

\begin{document}

\title{Sequential Multiple Structural Damage Detection and Localization: A Distributed Approach}

\author{Yizheng Liao, Ram Rajagopal \\ Emails:\textit{\{yzliao,ramr\}@stanford.edu}}

\maketitle

\begin{abstract}
As essential components of the modern urban system, the health conditions of civil structures are the foundation of urban system sustainability and need to be continuously monitored. In Structural Health Monitoring (SHM), many existing works will have limited performance in the sequential damage diagnosis process because 1) the damage events needs to be reported with short delay, 2) multiple damage locations have to be identified simultaneously, and 3) the computational complexity is intractable in large-scale wireless sensor networks (WSNs). To address these drawbacks, we propose a new damage identification approach that utilizes the time-series of damage sensitive features extracted from multiple sensors' measurements and the optimal change point detection theory to find damage occurrence time and identify the number of damage locations. As the existing change point detection methods require to centralize the sensor data, which is impracticable in many applications, we use the probabilistic graphical model to formulate WSNs and the targeting structure and propose a distributed algorithm for structural damage identification. Validation results show highly accurate damage identification in a shake table experiment and American Society of Civil Engineers benchmark structure. Also, we demonstrate that the detection delay is reduced significantly by utilizing multiple sensors' data.
\end{abstract}

%
\vspace{2pc}
%
%
%
%

\section{Introduction}
\label{sec:intro}

Infrastructural deterioration has become an expanding problem all over the world. According to the American Society of Civil Engineers (ASCE) 2017 Report Card For American's Infrastructure, 9\% bridges in the U.S. are structural deficient and these bridges carry over 188 million trips per day \cite{asce2017report}. In addition, 24\% of public school buildings are rated as fair or poor condition. Besides aging structures, New buildings and bridges may have structural health issues as well. For example, in September 2018, a cracked steel beam was found in the newly built San Francisco Transbay Transit Center, which was only opened to the public for one month. In order to prevent any failure or damage of the structures during daily operation or before being repaired or replaced, it is critical and necessary to continuously monitor any structural change that may adversely affect the performance and safety of deficient structures and provide quick assessment of the severity of damaged during daily operation and immediately after extreme events like earthquakes and hurricanes.

During the past several decades, Structural Health Monitoring (SHM) system has been proven to be an economical and reliable system to assess infrastructure health condition \cite{doebling1998summary, farrar2001vibration}. Among many structural damage detection methods, the vibration-based approaches have received significant attentions because the costs of accelerometers are low and the vibrational characteristics of the structure will be changed when damage changes the structure's physical properties. Methods and applications of vibration-based damage detection are summarized in \cite{doebling1998summary,farrar2001vibration,carden2004vibration,fan2011vibration}. Traditionally, the vibration-based damage detection methods use the collected structural responses to directly update or estimate the physical parameters \cite{fan2011vibration,zimmerman1992eigenstructure,fritzen1998damage,bernal2002load}. These model-based approaches can provide an accurate and easily interpolated physical parameters but have high computational complexity, which makes them infeasible for quick health assessment and processing massive data collected from a large-scale sensor network. In addition, many model-based approaches are insensitive to local or minor damage even with a dense sensor network \cite{yao2012autoregressive,morassi1993crack,salawu1997detection}. Furthermore, these physical models are vulnerable to measurement noises and environment conditions \cite{li2010modal,zhang2012investigation}.

In recent years, statistical pattern recognition  (SPR) techniques have been investigated to find more reliable and efficient damage detection algorithms. In SPR, the damage sensitive features (DSFs) are extracted from the acquired structural responses and designed to be sensitive to structural changes. Example of features include natural frequencies \cite{morassi1993crack,hackmann2014cyber,maity2005damage}, mode shapes \cite{lee2005neural}, wavelet coefficients \cite{kim2004damage,young2011use, shahsavari2017wavelet}, autoregressive model coefficients \cite{yao2012autoregressive,nair2006time,bornn2015modeling,zhang2007statistical}. After extracting damage sensitive features (DSFs), the damage is detected through changes or outliers in features rather than changes of the structural properties. As a result, the SPR-based damage detectors usually do not require the structural parameters as a prior. This advantage allows us to perform the analysis at sensor device directly as well as combine features from multiple sensors for a more comprehensive structural health assessment. The damage detection methods include hypothesis testing \cite{nair2006time,zhang2007statistical}, support vector machine \cite{oh2009damage}, neural network \cite{wu1992use,fang2005structural,yeung2005damage,abdeljaber2017real}, damage indication index \cite{yao2012autoregressive,nair2007time, gul2009statistical}, Gaussian process \cite{balafaswavelet}, and Bayesian network \cite{huang2017hierarchical,bornn2015modeling}.

For rapid structural health assessment, many of these existing approaches have limit performance. For example,  \cite{nair2006time}, \cite{zhang2007statistical}, and \cite{hui2017structural} require a long period of structural response to accurately detect damage. Also, many algorithms only focus on the damage detection of single location. However, in many damage cases, multiple locations can be damaged simultaneously and we need to identify these locations within a short period of time \cite{ruvcevskis2009vibration, yun2010parameter}. Furthermore, it becomes promising to deploy the large-scale wireless sensor networks for monitoring structural health condition due to many successful efforts to design and develop wireless sensor networks (WSNs) for SHM \cite{lynch2002wireless,wang2007wireless,nagayama2007structural,spencer2010wireless,liao2014snowfort}. These sensors can support continuously monitoring and provide accurate and massive data. But the existing approaches usually require to collect all data to a centralized device or a base station. In WSNs, the sensors are usually powered by batteries and the wireless communication consumes more power than the onboard computation \cite{liao2014snowfort}. Therefore, the frequent data collection will significantly reduce the lifetime of sensors. 

In order to tackle the challenges discussed above, we propose a sequential and distributed algorithm to detect and localize damage at multiple locations simultaneously. Specifically, we formulate the damage identification problem as a sequential hypothesis testing and apply the change point detection method \cite{tartakovsky2005general} to detect and localize damage. In the change point detection framework, the detector observes a sequence of DSFs and reports one or multiple damage events when it detects a change of DSFs probability distribution due to some events at an unknown time. The objective is identifying a damage event as quickly as possible subject to a fixed probability of false alarm. Unlike previous applications \cite{noh2013sequential,liao2018structural} of change point detection in SHM, we propose a set of detectors that not only quickly find the earliest damage location but also reports the number of damage locations. For these detectors, we prove the asymptotic bounds of expected detection delays. 

For the proposed detection rules, a key step is computing the marginal probabilities over multiple damage variables, which represent the structural components we want to monitor. As will be clear in the sequel, the computational complexity of the marginal probabilities will grow up exponentially with the number of damage variables. To address this computational challenge, we use a probabilistic graphical model to represent the target structure and the WSN that is used to collect DSFs. The graphical model not only visualizes the statistical dependence among the damage variables, which is usually unobservable, and the DSFs collected from sensors but also provides a computationally efficient algorithm for computing the marginal probabilities, in a sequential and distributed manner. With the graphical model formulation, the damage identification is performed at each sensor and does not need to centralize data.

The rest of the paper is organized as follows. Section~\ref{sec:single} introduces the sequential damage identification process, defines the DSF, and proposes the sequential method for single damage detection. In Section~\ref{sec:multiple}, we extend the algorithm for single damage detection to identify multiple damage locations. By using the graphical model to formulate the damage variables and DSFs, we propose to use the message-passing algorithm to compute the marginal probabilities in a distributed manner. Also, we propose a set of damage detectors that find the earliest damage location and report the number of damage locations. Additionally, we prove the asymptotic bounds of the average detection delays for these detectors. In Section~\ref{sec:validation}, we validate the proposed multiple damage identification algorithm using a shake table experiment data set and a simulation data set generated from ASCE benchmark structure. Section~\ref{sec:conclusion} draws summaries and conclusions.


\section{Sequential Damage Detection at Single Location}
\label{sec:single}
The process of damage detection and localization algorithm (see Fig.~\ref{fig:process}) consists of three major steps: (i) collection of structural responses, (ii) extraction of DSFs, and (iii) detection and localization of damage. In Step 1, the acceleration responses are acquired from multiple sensors sequentially. In Step 2, the DSFs are extracted from the collected signals. The DSF is required to be sensitive to damage patterns. In this paper, we us the coefficients of the acceleration autoregressive (AR) model as an example of DSF. \cite{nair2006time} and \cite{noh2013sequential} prove that the AR model coefficients extracted from the acceleration signals are related to structural parameters and sensitive to structural damage. We want to highlight that the proposed algorithm also works for other types of DSFs, such as wavelet coefficients \cite{shahsavari2017wavelet,liao2015sequential, nair2009derivation}, and AR model coefficients of angular velocity \cite{liao2016angular}. The DSF extraction includes two steps: (i) normalization and (ii) AR model fitting. For data normalization, the obtained responses are divided into chunk with size $N$ and then normalized before feature extraction, as shown below:
\[
\tilde{a}_i^n[t] = \frac{a_i^n[t] - \mu_i^n}{\sigma_i^n},
\]
where $a_i^n[t]$ denotes $t$-th sample in chunk $n$ collected at sensor $i$, $\mu_i^n$ and $\sigma_i^n$ denote the mean and the standard deviation of the $i$th chunk.

\putFig{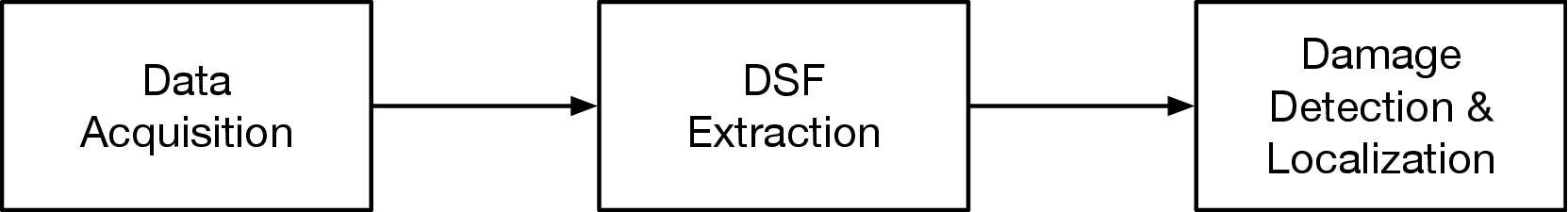}{Process of Sequential Damage Detection and Localization.}{0.7\linewidth}

After normalization, the normalized responses $\tilde{a}_i^n[t]$ are fitted with a single-variant auto-regressive (AR) model of order $p$,
\begin{equation}
	\tilde{a}_i^n[t] = \sum_{k=1}^p\theta_k[n] \tilde{a}_i^n[t-k] + \epsilon^n[t],
\end{equation} 
where $\theta_k$ is the $k$th AR coefficient and $\epsilon^n[t]$ is the residual. For connivance, we define $\mathbf{x}_i[n] \in \reals^m$ as the DSF of sensor $i$ extracted from the $n$-th chunk. $\mathbf{x}_i[n]$ is defined as a vector because it can contain multiple AR coefficients as DSF, e.g., $\mathbf{x}_i[n] = [\theta_1[n], \theta_2[n]]^T$. We sequentially collect structural responses. Therefore, we have a time-series of DSFs, e.g., $\mathbf{x}_i^N = \{\mathbf{x}_i[1], \mathbf{x}_i[2], \cdots, \mathbf{x}_i[N]\}$.

In Step 3 of the sequential damage detection and localization process, we apply a statistical test to diagnose DSFs sequentially. Since DSFs are time-series signals, one way to represent the data is using random variables. Therefore, we model DSFs at sensor $i$ as a random vector $\mathbf{X}_i$ and $\mathbf{x}_i[n]$ is the realization of $\mathbf{X}_i$ at time $n$. As shown in \cite{nair2007time}, the coefficients of the acceleration-based AR models follow Gaussian distribution. Therefore, we assume the DSF $\mathbf{X}_i$ follow a multivariate Gaussian distribution. The mean and covariance matrix of $\mathbf{X}_i$ change after a damage event. As a result, the probability distribution of $\mathbf{X}_i$ is different before and after damage occurrence. Let $\lambda_j$ denote the damage occurrence time of component $j$. We assume that $\mathbf{X}_i$ follows $\mathcal{N}(\mu^0_i, \Sigma^0_i)$ in the normal operation (i.e., $N < \lambda_j$), and a different Gaussian distribution $\mathcal{N}(\mu^1_i, \Sigma^1_i)$ after the damage event (i.e., $N \geq \lambda_k$). We use $g_i$ to denote the pre-damage distribution probability density function (PDF) and $f_i$ to denote the post-damage PDF. Finding the damage occurrence time $\lambda_j$ sequentially is equivalent to perform the following hypothesis test at each time $N$:
\begin{eqnarray*}
    \mbox{Pre-damage} &\quad & \mathcal{H}_0: \lambda_j > N, \\
    \mbox{Post-damage} &\quad & \mathcal{H}_1: \lambda_j \leq N.
\end{eqnarray*}
A well-known approach for solving this sequential hypothesis testing problem is the change point detection method \cite{tartakovsky2005general}. Usually, the damage occurrence time is unpredictable. Therefore, we assume $\lambda_j$ as a discrete random variable with a probability mass function $\pi_j(\lambda_j)$. Now, we can use a Bayesian approach to find $\lambda_j$. In this paper, we assume $\lambda_j$ follows a geometric distribution with a parameter $\rho_j$. The join distribution of $\lambda_j$ and $\mathbf{X}_j$ can be written as
\begin{equation}
\label{eq:single_dsf_likeli}
P(\lambda_j, \mathbf{X}_i) = \pi_j(\lambda_j)P(\mathbf{X}_i|\lambda_j).
\end{equation}
When $\lambda_j = n$, all the DSFs $\mathbf{X}_i$ obtained before time $n$ follow the distribution $g_i$ and all the DSFs obtained at and after time $n$ follow the distribution $f_i$. Therefore, the likelihood probability $P(\mathbf{X}_i|\lambda_j)$ above is expressed as follows:
\[
P(\mathbf{X}_i = \mathbf{x}_i^N|\lambda_j=n) = \prod_{k=1}^{n-1}g_i(\mathbf{x}_i[k])\prod_{k=n}^N f_i(\mathbf{x}_i[k])
\]
for $n=1,2,\cdots,N+1$. When $\lambda = N+1$, it refers to the damage has not occurred and the data follow the distribution $g_i$.

For monitoring structural damage condition, we usually install multiple sensors at different locations. When a damage event occurs, the DSFs extracted from multiple sensors will change. Therefore, we can use DSFs extracted from multiple sensors to diagnose the structural health condition. Assuming DSFs extracted from $M$ sensors are affected by the health condition of component $j$, (\ref{eq:single_dsf_likeli}) can be written as
\begin{equation}
    \label{eq:multiple_dsf_likeli}
    P(\lambda_j, \mathbf{X}_*) = \pi_j(\lambda_j)\prod_{i=1}^MP(\mathbf{X}_i|\lambda_j),
\end{equation}
where $\mathbf{X}_* = \{\mathbf{X}_1,\mathbf{X}_2,\cdots, \mathbf{X}_M\}$ and we assume DSFs extracted from each sensor are conditionally independent, given the damage condition. Now, finding the damage occurrence time $\lambda_j$ is equivalent to finding the post-damage posterior probability $P(\mathcal{H}_1|\mathbf{X}_*) = P(\lambda_j \leq N|\mathbf{X}_* = \mathbf{x}_*^N)$ at each time $N$, where $\mathbf{x}_*^N = \{\mathbf{x}_1^N, \mathbf{x}_2^N, \cdots, \mathbf{x}_M^N\}$ is the collection of all $M$ sensors' DSF observations. If the posterior probability is large enough, we can declare that component $j$ is damaged. At each time $N$,

\begin{eqnarray}
P(\lambda_j \leq N | \mathbf{x}_*^N) &=& \sum_{n=1}^N \frac{P(\lambda_j = n, \mathbf{x}_*^N)}{P(\mathbf{x}_*^N)}, \label{eq:single_post}\\
&=& \frac{1}{P(\mathbf{x}_*^N)}\sum_{n=1}^N\pi_j(\lambda_j = n)P(\mathbf{x}_*^N|\lambda_j = n), \\
&=& C\sum_{n=1}^N \pi_j(n) \prod_{i=1}^M P(\mathbf{x}_i^N|\lambda_j=n), \\
&=& C\sum_{n=1}^N \pi_j(n) \prod_{i=1}^M \left( \prod_{k=1}^{n-1}g_i(\mathbf{x}_i[k])\prod_{k=n}^{N}f_i(\mathbf{x}_i[k]) \right)
\end{eqnarray}
where $C$ is a normalization factor such that $\sum_{n=1}^{N+1}P(\lambda_j=n|\mathbf{x}_*^N) = 1$. During the normal operation, $f_i(\mathbf{x}_i[k])$ small for all sensors and $P(\lambda_j \leq N | \mathbf{x}_*^N)$ is small. Once the component $j$ is damaged at time $\lambda_j = k \leq N$, all data collected at $k \geq \lambda_j$ follow $f_i(\mathbf{x}_i[k])$ and $P(\lambda_j \leq N | \mathbf{x}_*^N)$ is large. Thus, we can set a threshold and declare an damage event when the posterior probability surpasses this threshold. 

\subsection{Optimal Damage Detection}
In the change point detection problem, there are two performance metrics: \textit{probability of false alarm} and \textit{expected detection delay}. The former metric is the probability that a detector falsely declares a damage event in the normal operation. If $\tau_j$ denotes the time to detect when the component $j$ is damaged, the probability of false alarm is defined as $P(\tau_j < \lambda_j)$. The latter metric describes the average latency that the damage event is detected after it has occurred. The expected detection delay is defined as $E(\tau_j - \lambda_j | \tau_j \geq \lambda_j)$. For structural damage detection, we want to find the damage time $\lambda_j$ as quickly as possible with a constraint of the maximum probability of false alarm $\alpha_{fa}$, i.e.,
\begin{eqnarray}
\mbox{minimize}_{\tau_j} &\quad & E(\tau_j - \lambda_j | \tau_j \geq \lambda_j) \label{eq:opt} \\
\mbox{subject to} &\quad & P(\tau_j < \lambda_j) \leq \alpha_{fa}. \nonumber
\end{eqnarray}
By the Shiryaev-Roberts-Pollaks procedure \cite{pollak2009optimality}, we have the following lemma to solve the optimization problem in (\ref{eq:opt}).

\begin{lemma}
\label{thm:decision_rule}
Given a maximum probability of false alarm $\alpha_{fa}$, the following detection rule
\begin{equation}
\label{eq:single_decision}
    \tau_j = \inf\{N \geq 1: P(\lambda_j \leq N|\mathbf{x}_*^N) \geq 1-\alpha_{fa}\},
\end{equation}
is asymptotically optimal \cite{tartakovsky2005general}.
\end{lemma}
With Lemma~\ref{thm:decision_rule}, the threshold for declaring the damage event is $1-\alpha_{fa}$. Lemma~\ref{thm:single_opt} shows the asymptotically optimal expected detection delay.

\begin{lemma}
\label{thm:single_opt}
For a given probability of false alarm $\alpha_{fa}$, the detection rule in (\ref{eq:opt}) achieves the asymptotically optimal detection delay
\begin{equation}
    D(\tau_j) = E(\tau_j - \lambda_j | \tau_j \geq \lambda_j) = \frac{|\log\alpha_{fa}|}{-\log(1-\rho_j) + \sum_{i=1}^M D_{KL}(f_i\|g_i)},
\end{equation}
as $\alpha_{fa} \rightarrow 0$, where $D_{KL}(f_i\|g_i)$ is the Kullback-Leibler distance \cite{amini2013sequential}.
\end{lemma}
The Kullback-Leibler (KL) distance is one of widely used metrics to describe the distance between two distributions. For probability density functions $g$ and $f$, the KL distance is defined as \cite{cover2012elements}
\begin{equation}
	D_{KL}(f\|g) = \int_{-\infty}^\infty f(x)\log\frac{f(x)}{g(x)}dx \geq 0.
\end{equation}
The equality holds when $f$ and $g$ are identical. Also, the KL distance is asymmetric, i.e., $D_{KL}(f\|g) \neq D_{KL}(g\|f)$. Since we assume DSFs follow multivariate Gaussian distribution, the KL distance can be computed as 
\[
    D_{KL}(f_i\|g_i) = \frac{1}{2}\left(tr\left((\Sigma_i^0)^{-1}\Sigma_i^1\right)+(\mu_i^0 - \mu_i^1)^T(\Sigma_i^0)^{-1}(\mu_i^0 - \mu_i^1) - R + \ln\left(\frac{\det\Sigma_i^0}{\det\Sigma_i^1}\right)\right),
\]
where $R$ denotes the number of elements in the DSF $\mathbf{X}_i$.

In Lemma~\ref{thm:single_opt}, if the damage variable parameter $\rho_j$ and the probability of false alarm $\alpha_{fa}$ are fixed, the detection delay is shorter if more sensors are installed to monitor the component $j$. Also, when the sensors are installed closer to the component $j$, the DSFs extracted from these sensors will change more significant after the damage event. Therefore, $D_{KL}(f_i\|g_i)$ is larger and the detection delay is shorter. 

In summary, when a new group of DSF $\mathbf{x}_*[n]$ is available, we compute the post-damage posterior probability according to (\ref{eq:single_post}) and then apply the optimal detection rule in (\ref{eq:single_decision}) to assess the structural health condition. Since we process the data sequentially, our method can provide real-time structural health assessment information for civil engineers and structure managers.


\section{Sequential Damage Detection at Multiple Locations}
\label{sec:multiple}
In the previous section, we propose an optimal detection rule to diagnose the health condition of a single component. However, buildings and bridges are composed of many components, such as braces, beams, and columns. Therefore, we need to monitor the health conditions of multiple components simultaneously. In this section, we will extend the proposed method for single location damage detection to identify multiple failure components in a structure.

With the prior knowledge, an expert can identify some critical components of a structure. Assuming we want to monitor $d$ components of a structure, similar to the previous section, we use $\lambda_j$ to denote the failure time of the component $j$ and $\pi_j(\lambda_j)$ to model the failure time distribution. Also, we use $\lambda_* = \{\lambda_1, \lambda_2, \cdots, \lambda_d\}$ to denote a collection of $d$ potential damage variables. Here, we assume each damage variable is independent and can occur simultaneously.

As discussed in the previous section, we can deploy a massive wireless sensor network in a structure today. Therefore, we can utilize DSFs extracted from these sensors to monitor the structural health condition. Specifically, given the damage variables $\lambda_*$ and DSFs from multiple sensors $\mathbf{X}_*$, we have the following joint distribution:
\begin{equation}
\label{eq:multi_joint}
    P(\lambda_*, \mathbf{X}_*) = \prod_{j=1}^d \pi_j(\lambda_j)\prod_{i=1}^M P(\mathbf{X}_i|\lambda_*).
\end{equation}
Then, one may compute the post-damage posterior probability $P(\lambda_*|\mathbf{x}_*^N)$ and apply the optimal detection rule in Lemma~\ref{thm:decision_rule} to diagnose each damage variable $\lambda_j$ to determine the health condition of the component $j$. This approach will work for small structures or a small size wireless sensor network and become infeasible for a structure with multiple critical components or a large-scale wireless sensor network. Specifically, to apply the detection rule in Lemma~\ref{thm:decision_rule}, we need to compute the marginal probability, e.g.,
\begin{equation}
    \label{eq:marg}
    P(\lambda_1|\mathbf{x}_*^N) = \sum_{n_2=1}^{N+1}\sum_{n_3=1}^{N+1}\cdots\sum_{n_d=1}^{N+1}P(\lambda_2 = n_2, \lambda_3 = n_3, \cdots, \lambda_d = n_d|\mathbf{x}_*^N).
\end{equation}
In the marginalization operation above, we need to sum over $(n+1)^{d-1}$ values, which makes this process inefficient and intractable for a large scale structure. 

Fortunately, we can use the physical property of damage events to tackle the computational difficulty of marginalization. Particularly, damage is usually a local event \cite{farrar1999statistical}. When a component fails, the features extracted from the sensors installed near the failure component will have statistical change. By utilizing the property, the DSFs from a sensor will only depend on a subset of damage variables. A widely used method to visualize these statistical dependencies is  a probabilistic graphical model \cite{wainwright2008graphical, jordan2004graphical}. In addition, the graphical model formulation provides an efficient algorithm for computing marginal and conditional probability, which addresses the computational challenge in (\ref{eq:marg}). A graphical model $\mathcal{G}=(\mathcal{V}, \mathcal{E})$ is usually characterized by a set of vertices $\mathcal{V}$ and a set of edges $\mathcal{E}$. In our model, the vertex set $\mathcal{V}$ includes both the damage variables $\lambda_*$ and the DSF variable $\mathbf{X}_*$. The edges are directional and specify the how the damage variables (unobservable) impact sensors' DSFs (observable). Since the component failure only impacts the sensors near the component, we can assign edges between the damage variable and the DSF variable based on the prior knowledge of structure. Also, we can learn this statistical dependency based on collected DSFs \cite{hastie2015statistical}. Fig.~\ref{fig:mp} demonstrates an example of the graphical model. In Fig.~\ref{fig:mp}, the circle nodes are the damage variables and the rectangular nodes are the DSF variables. The dash line arrows indicate the statistical dependencies. For example, the damage variable $\lambda_1$ impacts the DSFs of Sensor 1 ($\mathbf{X}_1$) and Sensor 2 ($\mathbf{X}_2$). The DSFs of Sensor 4 ($\mathbf{X}_4$) only depends on $\lambda_4$.

\putFig{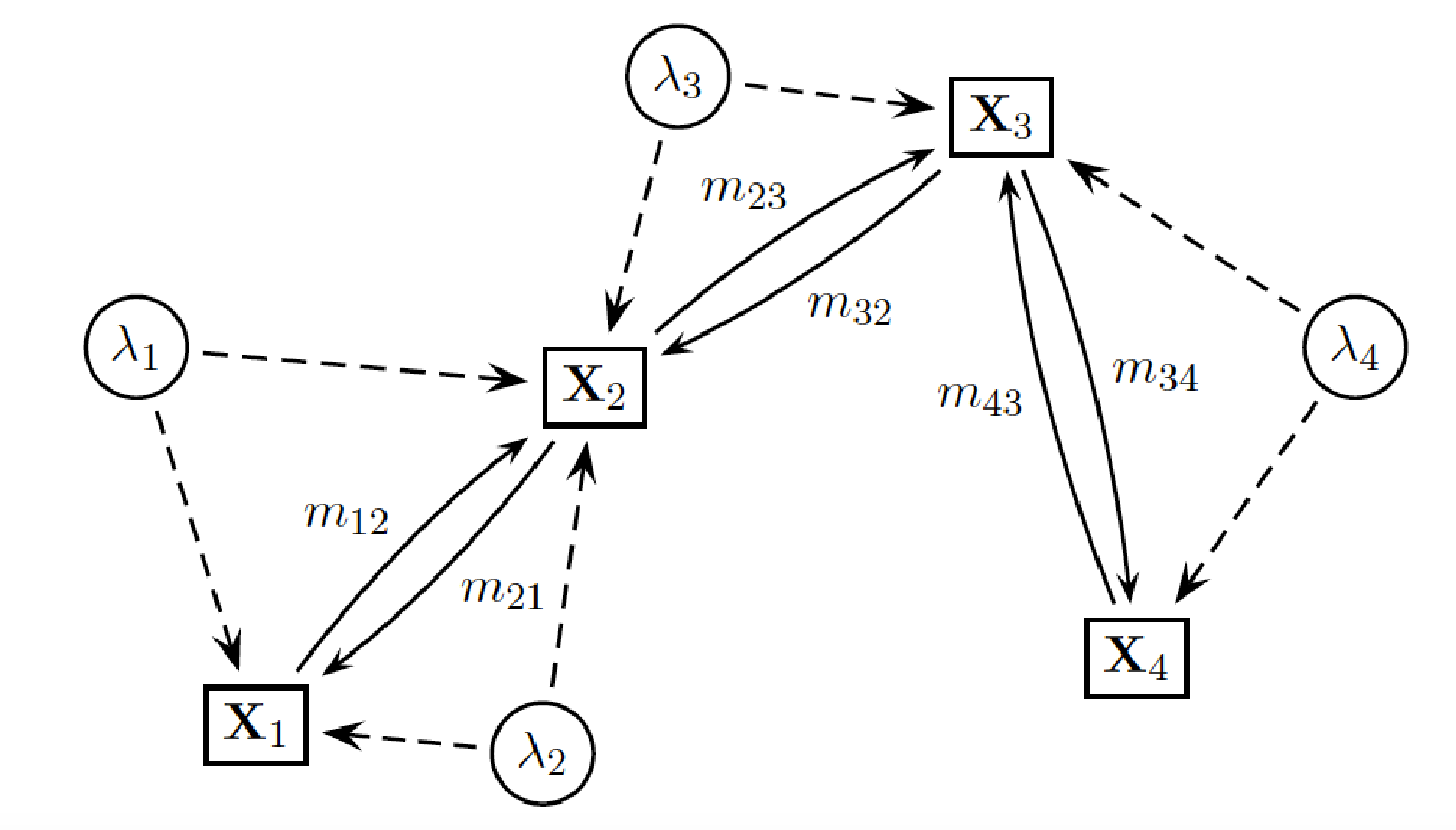}{Graphical model of the damage variables $\lambda_*$ and the DSF variables $\mathbf{X}_*$. The circle nodes are the damage variables. The rectangular nodes are the DSF variables. The dash line arrows indicate the statistical dependencies. The solid line arrows indicate the message passing paths.}{0.8\linewidth}

The graphical model has several advantages in SHM. First, it visualizes the statistical connection between sensors and the components we are monitoring. Also, it provides a view to interpret the structure from a statistical perspective, which enables us to understand the statistical inferences among sensors and apply different statistical methods for damage detection and localization. Furthermore, based on the graphical model, we can decompose a complex structure into several simple component and analyze each component independently. For example, unlike many previous works that declare damage for the whole structure, we can now examine each monitoring component based on partial observations, as shown in Fig.~\ref{fig:mp}. This decomposition also allows us to localize damaged components. At last, the graphical model serves as a tool to assistant the sensor deployment. 

Since the DSFs collected from a sensor only depends on a subset of damage variables, we can rewrite (\ref{eq:multi_joint}) as 
\begin{equation}
\label{eq:multi_joint2}
    P(\lambda_*, \mathbf{X}_*) = \prod_{j=1}^d \pi_j(\lambda_j)\prod_{i=1}^M P(\mathbf{X}_i|\lambda_{\mathcal{S}_i}),
\end{equation}
where $\mathcal{S}_i$ denotes the set that contains the indices of damage variables that affect $\mathbf{X}_i$. For the example in Fig.~\ref{fig:mp}, $\mathcal{S}_1 = \{1,2\}$ and $\mathcal{S}_3 = \{3, 4\}$. In (\ref{eq:multi_joint2}), $P(X_i|\lambda_{\mathcal{S}_i})$ depends on the status of multiple damage variables. For example, $\mathcal{S}_1 = \{1,2\}$ and therefore, we have three possible distributions, which is summarized in Table.~\ref{tab:damage_combin}

\begin{table}[htbp]
	\centering
	\caption{\label{tab:damage_combin} Summary of Possible Damage Distribution}
	\begin{tabular}{|c|c|}
	\hline
	$\lambda_1$ is triggered & $f_1^1$ \\
	\hline
	$\lambda_2$ is triggered & $f_1^2$ \\
	\hline
	$\lambda_1$ and $\lambda_2$ are triggered & $f_1^{\{1,2\}}$ \\
	\hline
	\end{tabular}	
\end{table}

From Table~\ref{tab:damage_combin}, we can see that the distribution after damage depend on the occurrence order of damage. Hence, given $\lambda_1 = n_1$ and $\lambda_2 = n_2$, the likelihood probability is defined as follows:
\begin{eqnarray}
	&& P(\mathbf{x}_1^N|\lambda_1=n_1,\lambda_2=n_2) \nonumber \\
	&=& \prod_{k=1}^{\min(n_1,n_2)-1}g_1(\mathbf{x}_1[k])\prod_{k=\min(n_1,n_2)}^{\max(n_1,n_2)-1}f_1(\mathbf{x}_1[k])\prod_{k=\max(n_1,n_2)}^N f_1^{\{1,2\}}(\mathbf{x}_1[k]), \label{eq:multi_cond} 
\end{eqnarray}
where $f_1(\mathbf{x}_1[k]) = f_1^1(\mathbf{x}_1[k])\indic{n_1 < n_2} + f_1^2(\mathbf{x}_1[k])\indic{n_1 > n_2}$ and $\indic{.}$ is the indicator function. The ranges of both $n_1$ and $n_2$ are from $1$ to $N+1$.  

In Section~\ref{sec:single}, we propose an optimal rule in Lemma~\ref{thm:decision_rule} to detect single damage variable. When there are multiple damage variables in the system, we can still apply this rule to identify if any of these faults has occurred. Besides detecting whether there is a damaged component in the structure, we are also interested in finding which component is damaged and the number of damaged components \cite{amini2013sequential}. To satisfy these two goals, we extend our proposed damage detection rule in (\ref{eq:single_decision}):
\begin{eqnarray}
\label{eq:min_rule}
	\mbox{\textit{minimum} rule} \qquad \tau_\mathcal{S}^{\min} &=& \inf\{N \geq 1:P(\lambda_\mathcal{S}^{\min} \leq N| \mathbf{x}_*^N) \geq 1 - \alpha_{fa}\},\\
\label{eq:max_rule}
\mbox{\textit{maximum} rule} \qquad	\tau_\mathcal{S}^{\max} &=& \inf\{N \geq 1:P(\lambda_\mathcal{S}^{\max} \leq N| \mathbf{x}_*^N) \geq 1 - \alpha_{fa}\},
\end{eqnarray}
where $\lambda_\mathcal{S}^{\min} = \min_{j \in \mathcal{S}}\lambda_j$ and $\lambda_\mathcal{S}^{\max} = \max_{j \in \mathcal{S}}\lambda_j$. The detection rule in (\ref{eq:min_rule}), which is referred as the \textit{minimum} rule, focuses on the detection of the earliest damage over the damage variable index set $\mathcal{S}$. Examples include the detection of single damage $\mathcal{S} = \{j\}$, the earlier one among two faults $\mathcal{S} = \{i,j\}$, and the earliest one among all damage variables. The detection rule in (\ref{eq:max_rule}), which is referred as the \textit{maximum} rule, focuses on how many components are damaged or whether all components in $\mathcal{S}$ have been damaged. In Section~\ref{sec:opt}, we will provide optimal bounds for these  detection rules.

Since we assign damage variables for the critical components, we can apply the \textit{minimum} rule to identify which damage variables have been triggered. If the damaged component is not associated with any damage variable, we can apply the \textit{maximum} rule to narrow down the potential area that contains the damaged components.

\subsection{Distributed Damage Detection: Message-passing Algorithm}
\label{sec:mp_alg}
Computing the posterior probability $P(\lambda_*|\mathbf{x}_*^N)$ usually requires centralizing the data from all sensors in the network. However, this approach is not desirable in practice. In wireless sensor networks, the wireless communication consumes significantly more power than the onboard computation \cite{liao2014snowfort}. Therefore, it is too costly to centralize data to perform the computation. Also, many networks do not even have a central base station. To minimize the energy consumption, we propose a distributed detection algorithm. As will be clear in the sequel, each sensor only needs to communicate with its neighbors. In this way, the computation of the posterior probability can be efficiently performed in a distributed manner at each sensor locally.


In the Bayesian statistics, the computation of posterior probability requires to compute the joint probability at first. Therefore, to propose our distributed algorithm, we need to decompose the joint probability in (\ref{eq:multi_joint2}) as follows:
\begin{equation}
	P(\mathbf{\lambda}_*,\mathbf{X}_*^n) = \prod_{j=1}^d \pi_j(\lambda_j)\prod_{i=1}^M P(\mathbf{X}_i|\lambda_{\mathcal{S}_i}) = \prod_{i=1}^M\alpha_i(\mathbf{\lambda}_{\mathcal{S}_i}),
\end{equation}
where $\alpha_i(\lambda_{\mathcal{S}_i})$ is a function that depends on the damage variables that affect DSF variable $X_i$. In our setup, $\alpha_i(\lambda_{\mathcal{S}_i})$ is either $P(X_i|\lambda_{\mathcal{S}_i})$ or $P(X_i|\lambda_{\mathcal{S}_i})\prod_{j: \mathcal{S} \subseteq \mathcal{S}_i}\pi_j(\lambda_j)$. Since the function $\alpha_i$ only utilizes DSFs extracted from sensor $i$, it can be computed at each sensor locally. We call $\alpha_i$ \textit{local kernel} and $\mathcal{S}_i$ \textit{local domain}. Table~\ref{tab:mp_local_kernel} shows the local kernels of the graphical model in Fig.~\ref{fig:mp}. We want to highlight that the local kernels are not unique.
\begin{table}[htbp]
    \centering
        \caption{\label{tab:mp_local_kernel} Local kernels of the graphical model in Fig.~\ref{fig:mp}.}
    \begin{tabular}{|c|c|c|}
    \hline
        Sensor & Local Domain ($\lambda_{\mathcal{S}_i})$) & Local Kernel $\alpha_i(\lambda_{\mathcal{S}_i})$  \\
    \hline
         1 & $\{\lambda_1, \lambda_2\}$ & $\pi_1(\lambda_1)P(\mathbf{X}_1|\lambda_1,\lambda_2)$ \\
         \hline
         2 & $\{\lambda_1, \lambda_2, \lambda_3\}$ & $\pi_2(\lambda_2)P(\mathbf{X}_2|\lambda_1,\lambda_2, \lambda_3)$ \\
         \hline
         3 & $\{\lambda_3, \lambda_4\}$ & $\pi_3(\lambda_3)P(\mathbf{X}_3|\lambda_3, \lambda_4)$ \\
         \hline
         4 & $\{\lambda_4\}$ & $\pi_4(\lambda_4)P(\mathbf{X}_4|\lambda_4)$ \\
    \hline
    \end{tabular}
\end{table}

In order to compute posterior probabilities, we could centralize the values computed by local kernels at each sensor and then normalize the product of local kernels. However, as we discussed above, this process is energy inefficient. If the graphical model is a tree, we can use the generalized distributive law (GDL) or the belief propagation algorithm to produce exact values of posterior probabilities \cite{amini2013sequential,wainwright2008graphical,aji2000generalized}. In the GDL algorithm, every sensor only needs to send a ``message'' to its neighbors. Specifically, if two sensors, or DSF variables, share any common damage variable, the message $m_{iq}^n$ from sensor $i$ to sensor $q$, at time $N$, is
\begin{equation}
\label{eq:msg}
	m_{iq}^N(\lambda_{\mathcal{S}_i \cap \mathcal{S}_q}) = \sum_{\lambda_{\mathcal{S}_i \backslash \mathcal{S}_q}} \alpha_i(\lambda_{\mathcal{S}_i}) \prod_{r \in \mathcal{N}(i) \backslash \{q\}}m_{ri}^N(\lambda_{\mathcal{S}_r \cap \mathcal{S}_i}),
\end{equation}
where $\mathcal{N}(i)$ denotes the index set of all neighbors that share at least one common damage variable with sensor $i$ and the set operator $\backslash$ refers to $\mathcal{S}_i \backslash \mathcal{S}_q = \{x \in \mathcal{S}_i, x \notin \mathcal{S}_q\}$. For example, in Fig.~\ref{fig:mp}, the neighbors of sensor $2$ are sensor $1$ and sensor $3$.

The message in (\ref{eq:msg}) is the product of the local kernel of sensor $i$ and all messages it has received from the neighbors of sensor $i$ other than sensor $q$ with filtering out the irrelative information (by marginalization). For example, in Fig.~\ref{fig:mp}, at time $N$, the message from sensor $3$ to sensor $2$ is
\begin{eqnarray*}
m_{32}^N(\lambda_{\mathcal{S}_3 \cap \mathcal{S}_2}) &=& m_{32}^N(\lambda_3) \\
&=& \sum_{\lambda_{\mathcal{S}_3 \backslash \mathcal{S}_2}}  \alpha_3(\lambda_{\mathcal{S}_3})\prod_{r \in \mathcal{N}(3)\backslash\{2\}} m_{r3}^N(\lambda_{\mathcal{S}_r \cap \mathcal{S}_3}) \\
&=& \sum_{\lambda_4}\pi_3(\lambda_3)P(\mathbf{X}_3|\lambda_3, \lambda_4)m_{43}^N(\lambda_4).
\end{eqnarray*}
Since the message is a function of the shared damage variables ($\lambda_{\mathcal{S}_i \cap \mathcal{S}_q}$), if two sensors do not share any common damage variable, there is no message between them. Therefore, every sensor only needs to communicate with its neighbors, as shown in Fig.~\ref{fig:mp}. This method is well known as the \textit{sum-product} algorithm \cite{wainwright2008graphical,jordan2004graphical}.

In order to communicate messages with all sensors, we pick up one sensor as the root. The root sensor starts to send messages to its neighbors. After receiving the message from the root, the root's neighbors send messages to their neighbors, excluding the root sensor. Since the graphical model is a tree, there is no loop. When all leaf sensors in the tree receive the messages, we repeat the process by sending messages from leaf sensors to their neighbors and then to the root. When the message passing is completed, on each sensor $i$, the joint probability $P(\lambda_{\mathcal{S}_i},\mathbf{X}_*)$ can be computed as follows:
\begin{equation}
\label{eq:beta}
	P(\lambda_{\mathcal{S}_i},\mathbf{X}_*) = \widetilde{\beta}_i(\lambda_{\mathcal{S}_i}) := \alpha_i(\lambda_{\mathcal{S}_i})\prod_{r \in \mathcal{N}(i)}m_{ri}^N(\lambda_{\mathcal{S}_r \cap \mathcal{S}_i}) \quad i = 1,2,\dots, M,
\end{equation}
where the $m_{ri}^N$ is the received message from the neighbors of sensor $i$ at time $N$. The joint probability is a product of the local kernel and all messages received from its neighbors. In this step, we do not exclude any message. Then, we normalize $\widetilde{\beta}_i(\lambda_{\mathcal{S}_i})$, i.e.,
\[
    \beta_i(\lambda_{\mathcal{S}_i}) = \frac{ \widetilde{\beta}_i(\lambda_{\mathcal{S}_i})}{\sum_{\lambda_{\mathcal{S}_i}=1}^{N+1}\widetilde{\beta}_i(\lambda_{\mathcal{S}_i})}.
\]
After normalization, $\beta_i(\lambda_{\mathcal{S}_i})$ is the posterior probability, $P(\lambda_{\mathcal{S}_i}|\mathbf{x}^N_*)$. For the graphical model in Fig.~\ref{fig:mp}, $\beta_1(\lambda_{\mathcal{S}_1}) = P(\lambda_1,\lambda_2|\mathbf{x}^N_*)$ and $\beta_2(\lambda_{\mathcal{S}_2}) = P(\lambda_1,\lambda_2,\lambda_3,\lambda_4|\mathbf{x}^N_*)$. We can see that $\beta_1(\lambda_{\mathcal{S}_1})$ is the marginalized probability of $ \beta_2(\lambda_{\mathcal{S}_2})$, i.e., $\sum_{\lambda_3,\lambda_4}\beta_2(\lambda_{\mathcal{S}_2}) = \beta_1(\lambda_{\mathcal{S}_1})$. Therefore, the detection rules can be applied at each sensor or at the sensor that contains all damage variables. In all the computations above, at time $N$, the range of $\lambda$ is from $1$ to $N+1$. When $\lambda \leq N$, the probability that the damage has occurred is computed. When $\lambda = N+1$, we compute the probability that the damage will happen in the future.

\putFig{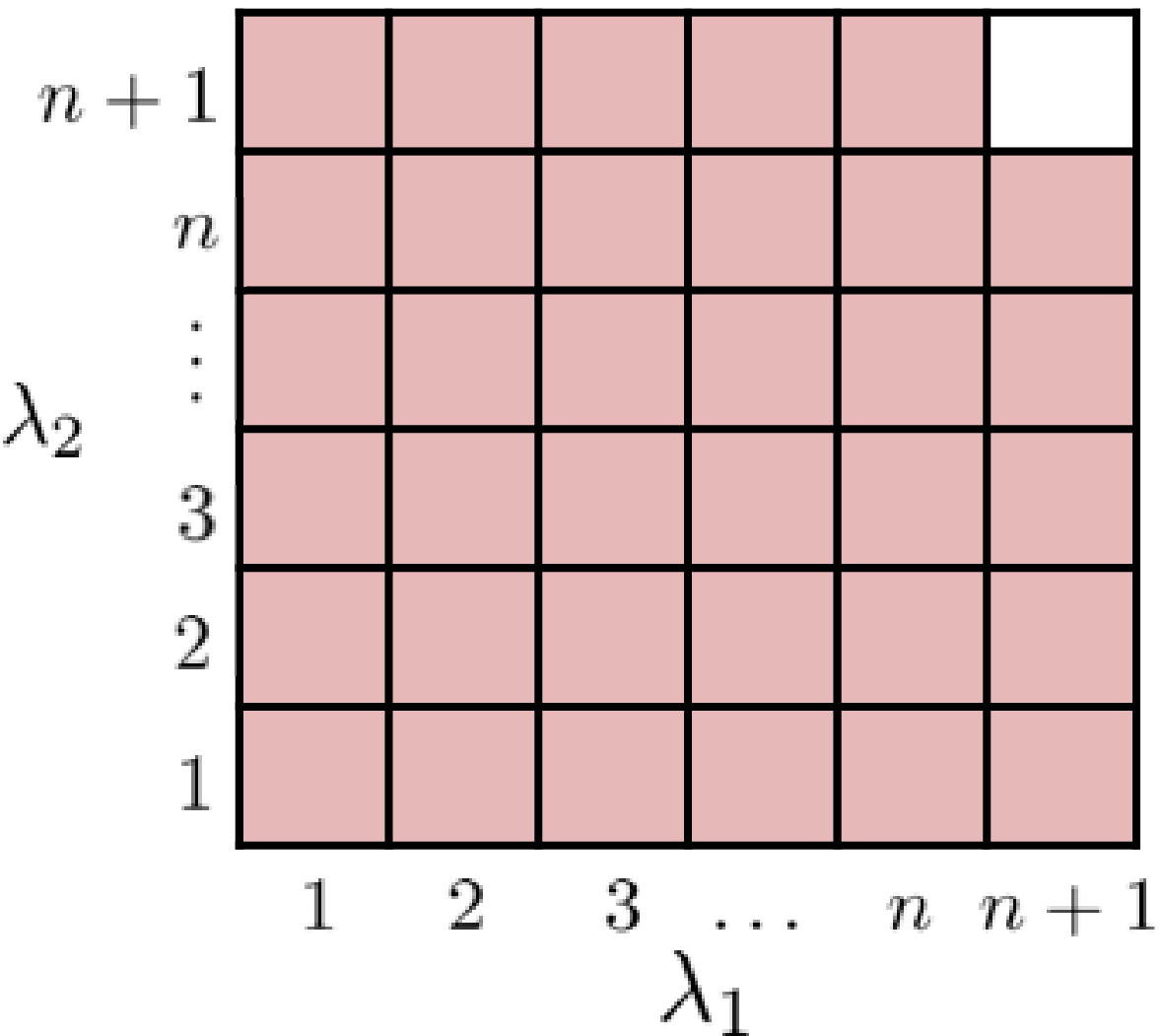}{The minimum rule with $\lambda_1$ and $\lambda_2$. $P(\min \{\mathbf{\lambda}_{\mathcal{S}_i}\} \leq n|\mathbf{x}_*^n)$ is the summation of the red areas.}{0.5\linewidth}

\putFig{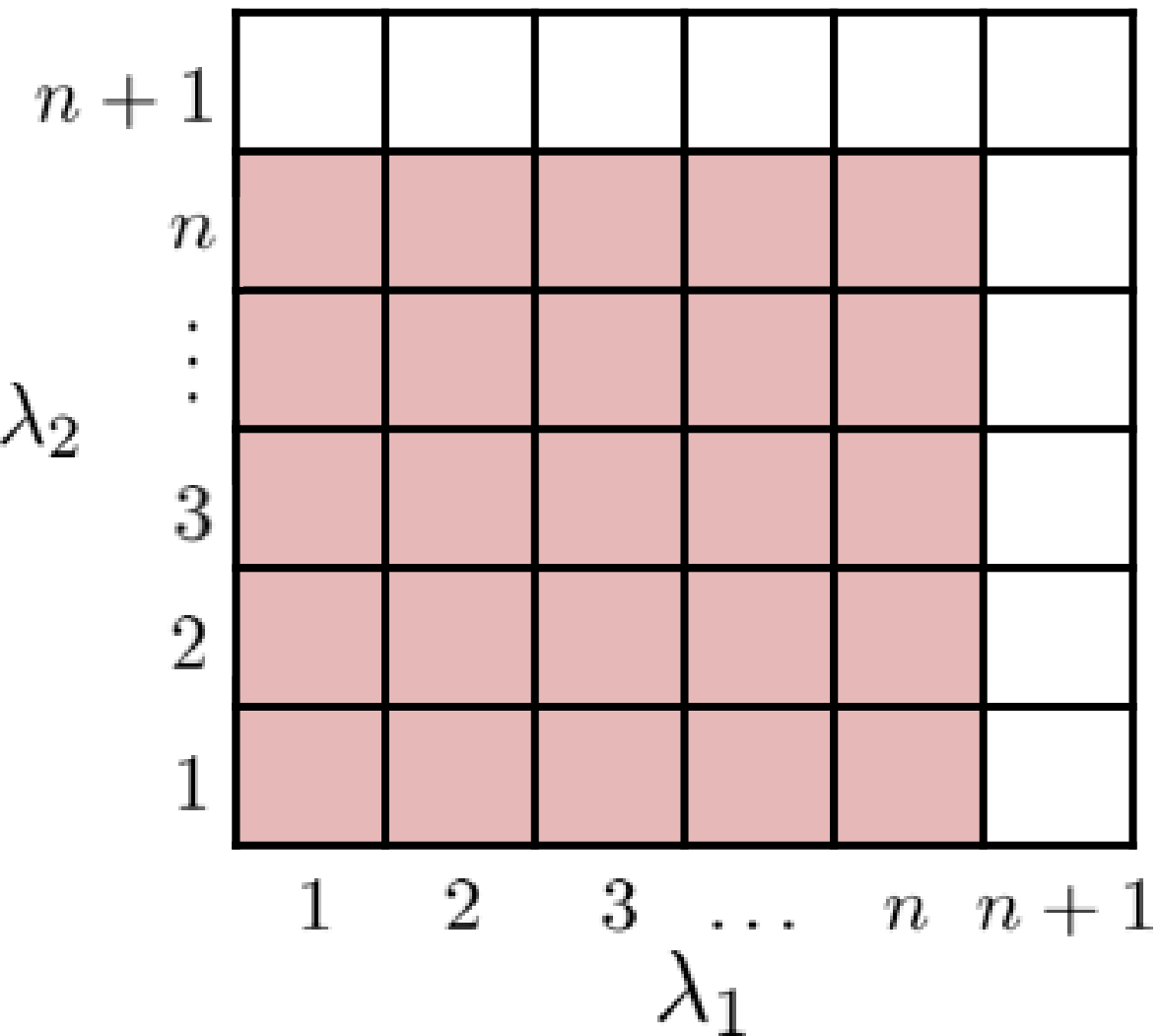}{The maximum rule with $\lambda_1$ and $\lambda_2$. $P(\max\{\mathbf{\lambda}_{\mathcal{S}_i}\} \leq n|\mathbf{x}_*^n)$ is the summation of the red areas.}{0.5\linewidth}

%
After getting $\beta_i(\lambda_{\mathcal{S}_i})$, we can apply the proposed detection rules in (\ref{eq:min_rule}) and (\ref{eq:max_rule}) to diagnose damage. Some examples of the detection rules are given below. Suppose sensor $i$ has statistical dependence with $\lambda_{\mathcal{S}_i} = \{\lambda_1,\lambda_2,\dots,\lambda_d\}$, 
\begin{itemize}
	\item for detecting the earliest damage, i.e., $\min \{\mathbf{\lambda}_{\mathcal{S}_i}\}$, the posterior probability is
		\begin{equation}
		\label{eq:post_min}
			P( \min \{\mathbf{\lambda}_{\mathcal{S}_i}\} \leq N|\mathbf{x}_*^N) = 1 - \beta_i(\mathbf{\lambda}_{\mathcal{S}_i} = N+1).
		\end{equation}
		This posterior probability can be visualized in Fig.~\ref{fig:min_rule} for involving two damage variables;
		
	\item for detecting all potential damage locations, i.e., $\max\{\mathbf{\lambda}_{\mathcal{S}_i}\}$, the posterior probability is
		\begin{equation}
		\label{eq:post_max}
			P(\max\{\mathbf{\lambda}_{\mathcal{S}_i}\} \leq N|\mathbf{x}_*^N) = \sum_{\lambda_1 = 1}^N\dots\sum_{\lambda_d=1}^N \beta_i(\mathbf{\lambda}_{\mathcal{S}_i}).
		\end{equation}
		This posterior probability can be visualized in Fig.~\ref{fig:max_rule} for involving two damage variables.
		
	\item for detecting a single damage, e.g., $\lambda_1$, the posterior probability is
		\begin{equation}
		\label{eq:post_one}
			P(\lambda_1 \leq N | \mathbf{x}_*^N) = \sum_{\lambda_1=1}^N\sum_{\lambda_2=1}^{N+1}\dots\sum_{\lambda_d=1}^{N+1}\beta_i(\mathbf{\lambda}_{\mathcal{S}_i});
		\end{equation}
		The damage variables $\lambda_2,\lambda_3,\dots,\lambda_d$ are marginalized. Since $\lambda_{\mathcal{S}_i}$ is a subset of $\lambda_*$, this computation is more efficient.
	\item for detecting the earliest damage among a subset, e.g. $\min\{\lambda_1,\lambda_2\}$, the posterior probability is
		\begin{equation}
		\label{eq:post_minS}
			P(\min\{\lambda_1,\lambda_2\} \leq N  | \mathbf{x}_*^N) = 1-\sum_{\lambda_3=1}^{N+1}\dots\sum_{\lambda_d=1}^{n+1} \beta_i(\mathbf{\lambda}_{\mathcal{S}_i});
		\end{equation}
	\item for detecting all faults over a subset of damage variables, e.g. $\max\{\lambda_1,\lambda_2\}$, the posterior probability is 
		\begin{equation}
		\label{eq:post_maxS}
			P(\max\{\lambda_1,\lambda_2\} \leq N | \mathbf{x}_*^N) = 
				\sum_{\lambda_1=1}^N\sum_{\lambda_2=1}^{N}\sum_{\lambda_3=1}^{N+1}\dots\sum_{\lambda_d=1}^{N+1}\beta_i(\mathbf{\lambda}_{\mathcal{S}_i}).
		\end{equation}
		
\end{itemize}

Algorithm~\ref{alg:mp} summarizes the implementation of the proposed  message-passing algorithm and damage detection rules.

\begin{algorithm}[ht!]
	\caption{Distributed damage detection algorithm at time $N$}
	\label{alg:mp}	
	\centering
	\begin{algorithmic}[1]
	\State Choose one sensor in the network as the root of the tree.
	\State Initialize messages $m_{iq}^N$ to the all ones for all edges. Compute local kernel $\alpha_i(\mathbf{\lambda}_{\mathcal{S}_i})$ for each sensor $i$.
	\State Compute and pass messages $m_{iq}^N$ from sensor $i$ to sensor $q$ according to (\ref{eq:msg}). Send messages from the root to its children. Continue the process from the top of the tree to the bottom till all the leaf nodes are reached. 
	\State Repeat Step 3 but start from the leaf nodes. Send messages from the leaf nodes to their parents. Continue the process from the bottom of the tree to the top till the root sensor is reached. When compute $m_{iq}^n$ in (\ref{eq:msg}), use the messages received in Step 3.
	\State Compute $\widetilde{\beta}_i(\mathbf{\lambda}_{\mathcal{S}_i})$ based on (\ref{eq:beta}) for $\lambda_j = 1,2,\dots,N,N+1$ and all $j$ in $\mathcal{S}_i$. Then normalize $\widetilde{\beta}_i(\mathbf{\lambda}_{\mathcal{S}_i})$ to have $\beta_i(\mathbf{\lambda}_{\mathcal{S}_i})$ such that $\sum \beta_i(\mathbf{\lambda}_{\mathcal{S}_i}) = 1$.
	\State Compute the posterior probability according to one of the rules in (\ref{eq:post_min})-(\ref{eq:post_maxS}).
	\If{$P(\mathbf{\lambda}_{\mathcal{S}} \leq N|\mathbf{x}_*^N) \geq 1-\alpha_{fa}$}
		\State Declare damage occurrence and stop.
	\Else
		\State Repeat Step 2-6 with the DSFs extracted at time $N+1$
	\EndIf
	\end{algorithmic}

\end{algorithm}

\subsection{Asymptotic Optimality of Multiple Damage Detection Rules}
\label{sec:opt}
In Lemma~\ref{thm:single_opt}, we prove that the decision rule for the single damage variable detection follows the Shiryaev-Roberts-Pollak procedure and is optimal. The \textit{minimum} and \textit{maximum} rules also follow the Shiryaev-Roberts-Pollak procedure. Therefore, the asymptotic detection delay has the following format:
\begin{equation}
\label{eq:opt_delay}
	D(\tau_\mathcal{S}) = E(\tau_\mathcal{S} - \lambda_\mathcal{S}|\tau_\mathcal{S} \geq \lambda_\mathcal{S}) = \frac{|\log\alpha_{fa}|}{q_{\lambda_\mathcal{S}}+I_{\lambda_\mathcal{S}}},
\end{equation}
where $q_{\lambda_\mathcal{S}}$ is a function of the prior distribution and $I_{\lambda_\mathcal{S}}$ is a function of the KL distances. As shown in Fig.~\ref{fig:mp}, each damage variable $\lambda_j$ has statistical links with the multiple sensors. Therefore, if $\lambda_j$ is triggered, the data distributions at multiple sensors will be changed. In \cite{amini2013sequential}, the authors have proven that for the \textit{minimum} rule, $I_{\lambda_\mathcal{S}}$ can be expressed as the summation of the KL distances at multiple sensors. In this paper, we extend the results to the \textit{maximum} rule. Let us highlight some particular cases of interest:
\begin{enumerate}
	\item To detect a single fault $\lambda_j$, we can use the \textit{minimum} rule, i.e., $\min\{\lambda_j\}$, the asymptotically optimal delay is
		\begin{equation}
		\label{eq:opt_single}
		\frac{|\log\alpha_{fa}|}{-\log(1-\rho_j) + \sum_{i \in \mathcal{Q}_j} D_{KL}(f^j_i\|g_i)},
		\end{equation}
		where $\mathcal{Q}_j$ contains the sensors that have statistical dependence with $\lambda_j$. For example, in Fig.~\ref{fig:mp}, $\mathcal{Q}_1 = \mathcal{Q}_2 = \{1,2\}$ and $\mathcal{Q}_3 = \{2,3\}$. $D_{KL}(f^j_i\|g_i)$ denotes the KL distance of $\lambda_j$ at sensor $i$. When the entire network only contains one damage variable, such as the single damage detection in Section~\ref{sec:single}, $D_{KL}(f^j_i\|g_i) = D_{KL}(f_i\|g_i)$. However, if there exists multiple damage variables, such as the example in (\ref{eq:multi_cond}), the post-damage distribution is complex. Therefore, we only consider the case that $\lambda_j$ is triggered but the rest damage variables have not been triggered. Hence, for the example in (\ref{eq:multi_cond}), we only use $D_{KL}(f_1^1\|g_1)$ to compute the optimal bound.
	\item To detect the earliest fault among multiple damage variables, i.e., $\min\{\lambda_\mathcal{S}\}$, the asymptotically optimal delay is 
		\begin{equation}
		\label{eq:opt_min}
		\frac{|\log\alpha_{fa}|}{-\sum_{j \in \mathcal{S}}\log(1-\rho_j) + \sum_{j \in \mathcal{S}}\sum_{i \in \mathcal{Q}_j} D_{KL}(f^j_i\|g_i)}.
		\end{equation}
		Since we focus on multiple damage variables, we summarize the prior information and KL distances of all nodes that are related to these damage variables. When we compute the optimal bound in a system with multiple damage variables, we also assume that only $\lambda_j$ is triggered but the rest damage variables remain silent. If $\lambda_\mathcal{S}$ only contains one damage variable, (\ref{eq:opt_min}) and (\ref{eq:opt_single}) are identical. 
	\item To detect all faults over a set of multiple damage variables, i.e., $\max\{\lambda_\mathcal{S}\}$, the asymptotically optimal delay is
		\begin{equation}
		\label{eq:opt_max}
		\frac{|\log\alpha_{fa}|}{-\sum_{j \in \mathcal{S}}\log(1-\rho_j) + \sum_{i: \mathcal{S} \subset \mathcal{S}_j}D_{KL}(f_j^{\mathcal{S}}\|g_j)},
		\end{equation}
		where $\mathcal{S}_k$ represents that an index set of damage variables that affect the measurements at sensor $k$. The probability density function (PDF) $f_k^{\mathcal{S}}$ describes the distribution after all damage variables in $\mathcal{S}$ have been triggered. For the example in (\ref{eq:multi_cond}), if we want to detect both $\lambda_1$ and $\lambda_2$, $I_1^{\{1,2\}} = D_{KL}(f_j^{\{1,2\}}\|g_j)$. 
\end{enumerate}
From the examples above, we can observe that the \textit{minimum} rule (\ref{eq:opt_min}) usually has lower detection delay than the \textit{maximum} rule (\ref{eq:opt_max}) because the \textit{minimum} rule uses more information. In Fig.~\ref{fig:min_rule} and Fig.~\ref{fig:max_rule}, the red area of the \textit{minimum} rule is larger than that of the \textit{maximum} rule. Therefore, the\textit{minimum} rule has lower delay than the \textit{maximum} rule. Also, utilizing all the data in the network can reduce the detection delay. To see this more clearly, if we only use the DSFs from sensor $i$ and ignore the share information, the stopping rule is $\tau_{loc} = \inf\{n: P(\lambda_{j} \leq N | \mathbf{x}^N_i) \geq 1 - \alpha_{fa}\}$, which only depends on local information for $\lambda_j \in \mathcal{S}_i$. The asymptotically optimal delay becomes $(-\log(1-\rho_j) + D_{KL}(f_j^i\|g_j))^{-1}|\log\alpha_{fa}|$. With the lack of information from other sensors, this delay is larger than the one in (\ref{eq:opt_single}). This observation is validated by simulations in Section~\ref{sec:validation}, where $\tau_{loc}$ is referred as the LOCAL rule and the rules in (\ref{eq:min_rule}) and (\ref{eq:max_rule}) are referred as the MP rule.

\section{Experimental and Numerical Results}
\label{sec:validation}
In this section, we apply the proposed damage identification algorithm to a data set collected from an indoor shake table experiment. Then, we use simulation data collected from the ASCE benchmark structure to validate the consistency and robustness of the proposed algorithm in the case where multiple faults occur simultaneously. We show that using the DSFs from all sensors to detect damage has a lower detection delay than only using local information.

\subsection{Indoor Shake Table Experiment Validation}
\label{sec:exp_data}

\subsubsection{Description of Experiment}
\label{sec:des_exp}
To validate the proposed damage detection algorithm, we use the experimental data obtained from a shake table experiment conducted at the National Center for Research on Earthquake Engineering (NCREE) in Taipei, Taiwan. In this experiment, two identical three-story steel frames were placed side-by-side on the same shake table, as shown in Fig.~\ref{fig:taiwan1}. The white noise excitation was applied in the North-South direction with $0.05g$ amplitude. 

\putFig{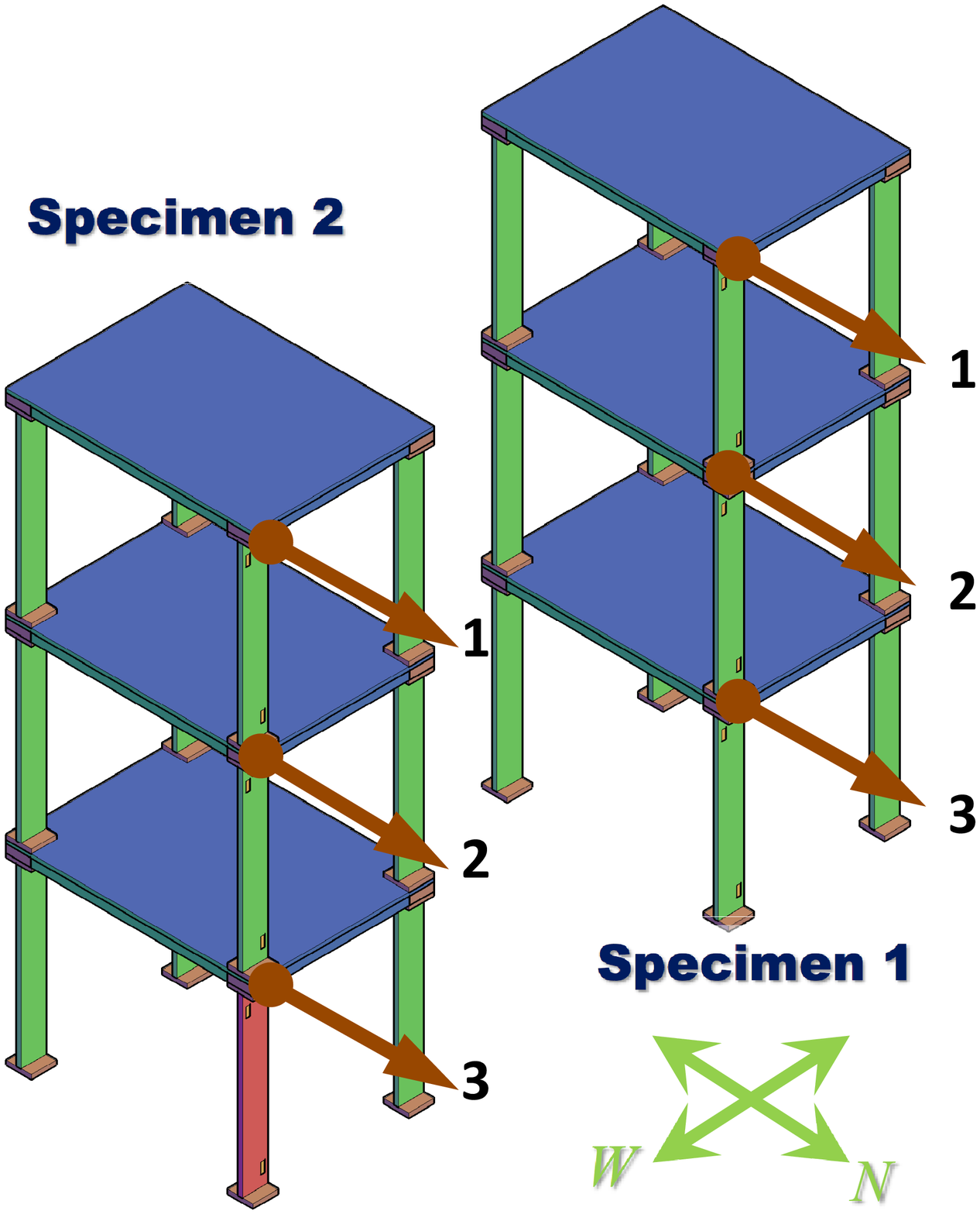}{The diagram of structures and sensors. The red column is the weakened column.}{0.5\linewidth}

This experimental data set contains both major and minor damage patterns. For the major damage pattern, we replace the North West column of Specimen 2 with a weakened column that had only $60\%$ thickness of the normal columns (see Fig.~\ref{fig:taiwan1}). The DSFs extract from Specimen 1 represent the pre-damage state and the features extract from Specimen 2 represent the post-damage state.
The minor damage pattern is introduced by the strong motions. The record data from the $1999$ Chi-Chi earthquake Station TCU $071$ is used as the base excitation of the experiment and is applied in the x-direction with amplitudes progressively increasing from $0.1g$ to $1.45g$. When the amplitude of the strong motion is above $0.85g$, the weakened column of Specimen 2 has the yielding effect, as indicated in Fig.~\ref{fig:yielding}.

\putFig{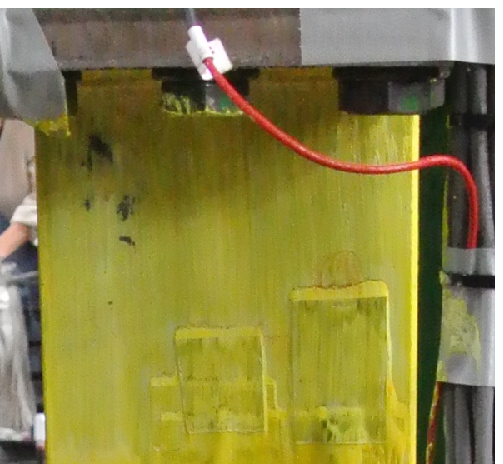}{Weakened column of Specimen 2 before any strong motion.}{0.8\linewidth}
\putFig{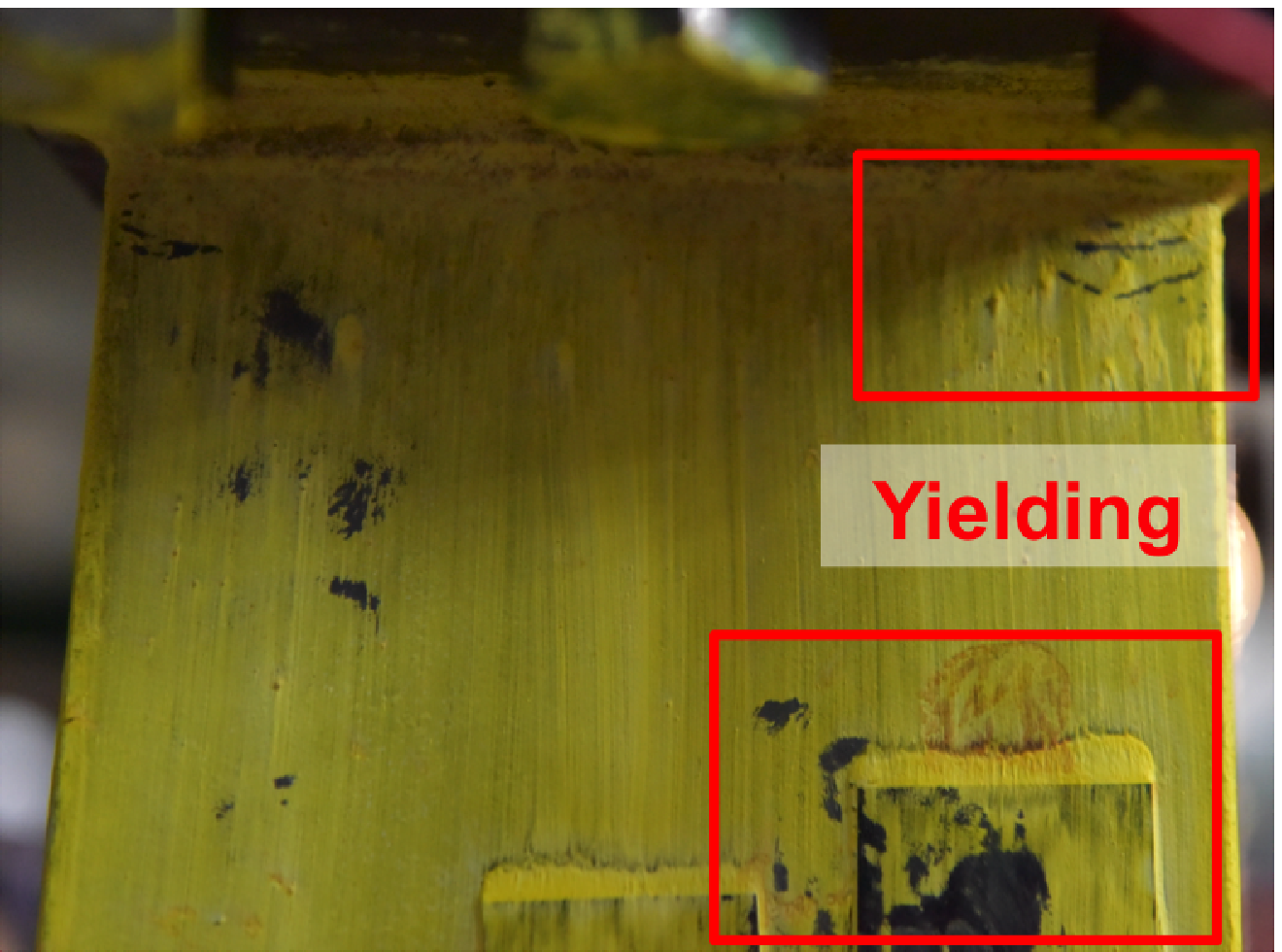}{Weakened column of Specimen 2 after the strong motion with an amplitude of $0.85g$.}{0.8\linewidth}


The one-dimensional accelerometers are installed for data acquisition as shown in Fig.~\ref{fig:taiwan1}. The measurements are acquired with a sampling frequency of $200$Hz and filtered by an anti-aliasing filter with a cut-off frequency of $50$Hz. The data are processed with the same procedure discussed in Section.~\ref{sec:single}. The normalized data are modeled as an AR model with an order of $p=7$, which is selected by using AIC values. The first AR coefficient $\theta_1$ is used as the DSF.

For the major damage pattern, we concentrate $40$ DSFs extracted from Specimen 1, which represent the pre-damage signal, and $40$ DSFs extracted from Specimen 2, which represent the post-damage signal. Hence, the true damage time is $41$. We assign a damage variable $\lambda_j$ for each floor and assume that the damage will affect the measurements of the sensors on the current floor and all floors above. Since we want to declare the damage based on the data rather than the prior information, we assume $\lambda_j$ follows a geometric distribution with $\rho = 0.001$. Table~\ref{tab:local_domain_tw} summarizes the local domains and kernels information and Fig.~\ref{fig:gp_tw} visualizes the graphical model.

\begin{table}[h!]
\caption{Local domains and local kernels of the experimental structure}
\label{tab:local_domain_tw}
\centering
\begin{tabular}{c| l |l}
\hline
Sensor & Local Domain $\lambda_{\mathcal{S}_j}$ & Local Kernel $\alpha_i(\lambda_{\mathcal{S}_j})$ \\
\hline
1st Floor & $\left\{\lambda_1\right\}$ & $\pi_1(\lambda_1)P(\mathbf{X}_1^n|\lambda_1)$ \\
\hline
2nd Floor & $\left\{\lambda_1,\lambda_2\right\}$ & $\pi_2(\lambda_2)P(\mathbf{X}_2^n|\lambda_1,\lambda_2)$ \\
\hline
3rd Floor & $\left\{\lambda_1,\lambda_2,\lambda_3\right\}$ & $\pi_3(\lambda_3)P(\mathbf{X}_{3}^n|\lambda_1,\lambda_2,\lambda_3)$  \\
\hline
\end{tabular}
\end{table}

\putFig{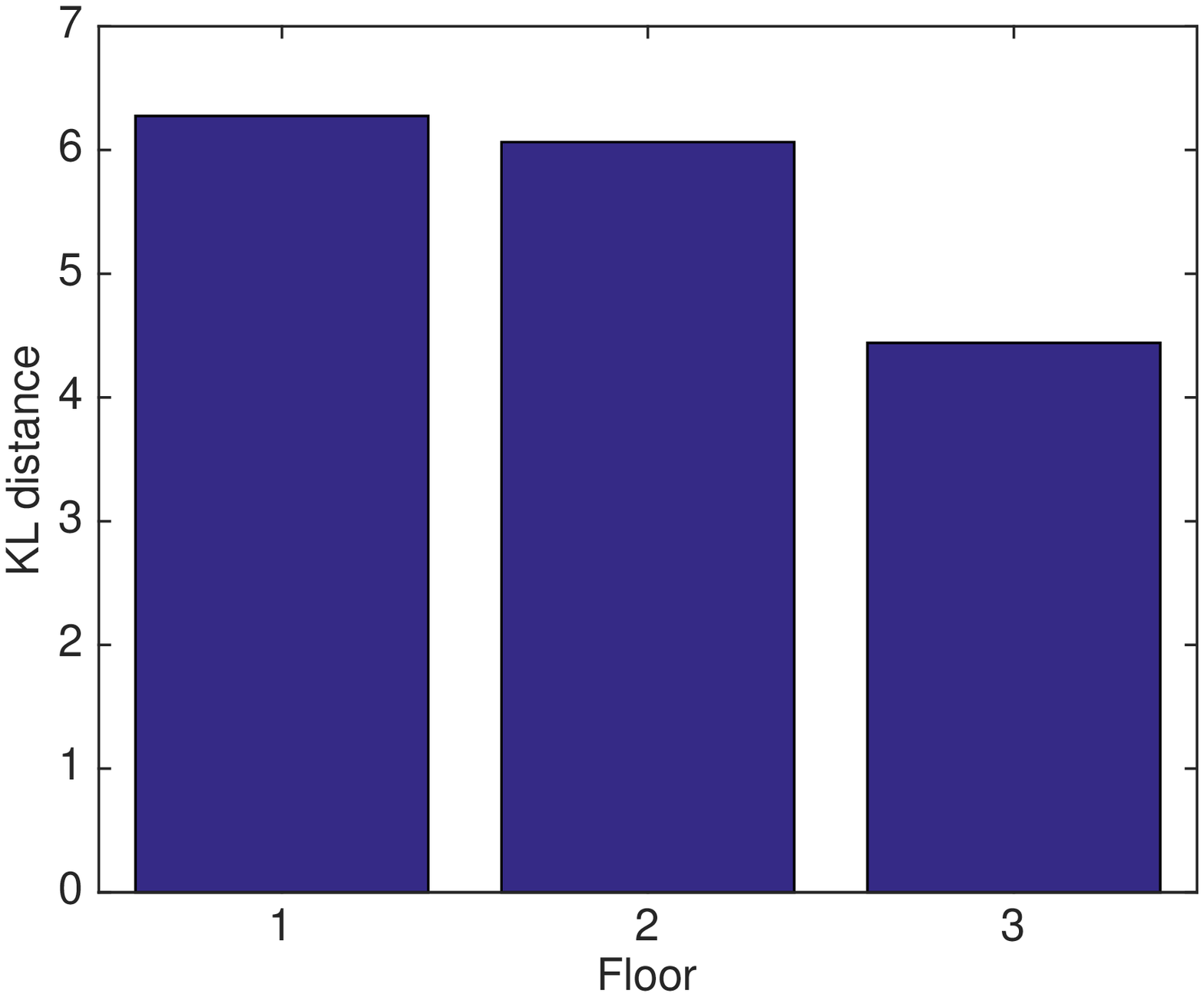}{KL distances between the pre-damage and post-damage DSF distributions at each floor for the major damage.}{0.8\linewidth}
\putFig{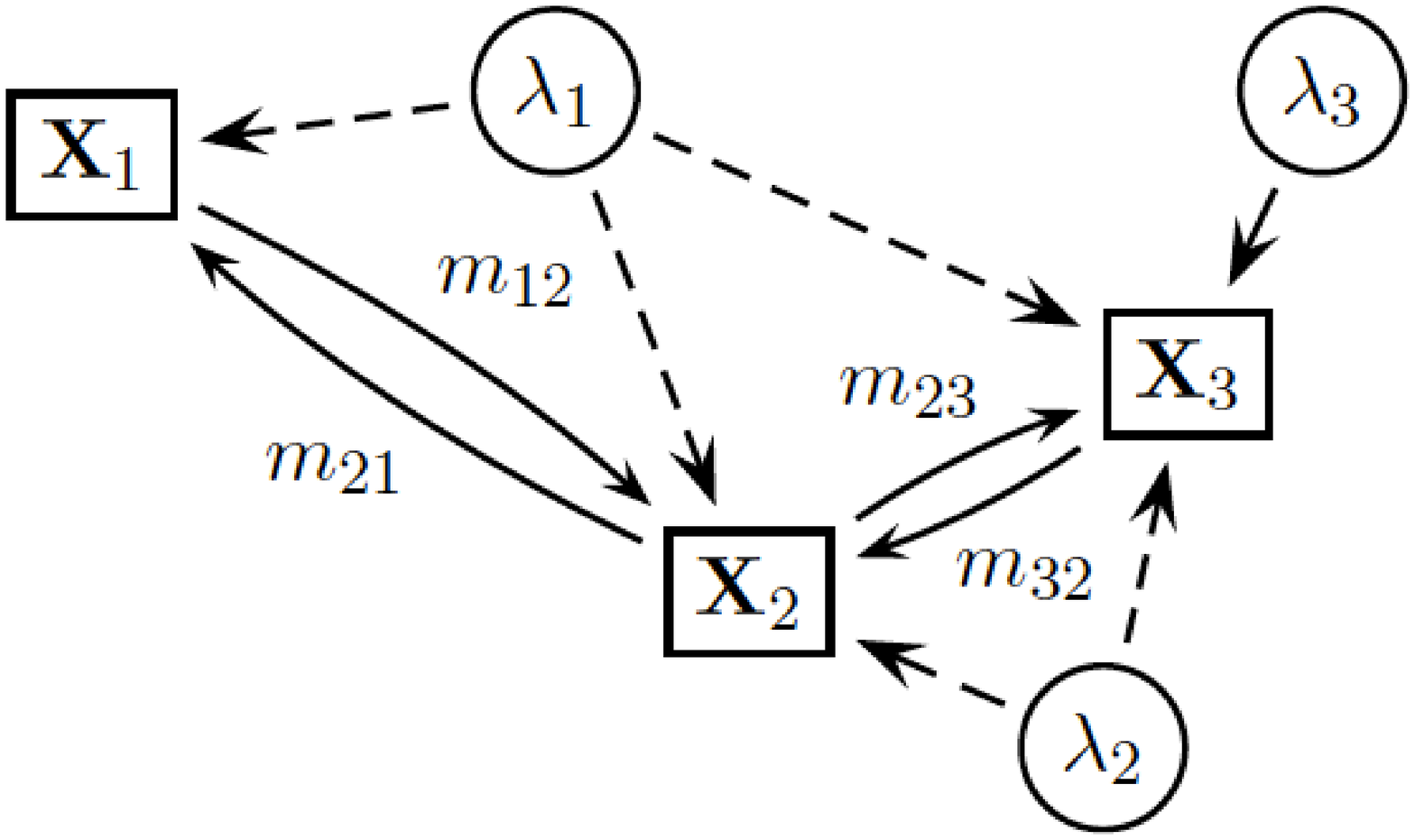}{Graphical model of damage variables ($\lambda$) and observable variables ($\mathbf{X}$) in the experimental structure. The dash arrows show the statistical dependencies and the solid arrows indicate the message passings.}{0.8\linewidth}


\subsubsection{Results and Discussions}
In this section, we apply the proposed algorithm to identify the damage on the first floor. We use the first AR coefficient $\theta_1$ as the DSF. We implement the \textit{minimum} rule to detect $\lambda_1$, i.e., $\min(\lambda_1)$. Fig.~\ref{fig:exp_major_KL} shows the KL distances of sensors at each floor. When the first floor is damaged, all sensors above the first floor have changes of their DSF distributions. This observation is consistent with our local domain assignment in Table.~\ref{tab:local_domain_tw}.

\putFig{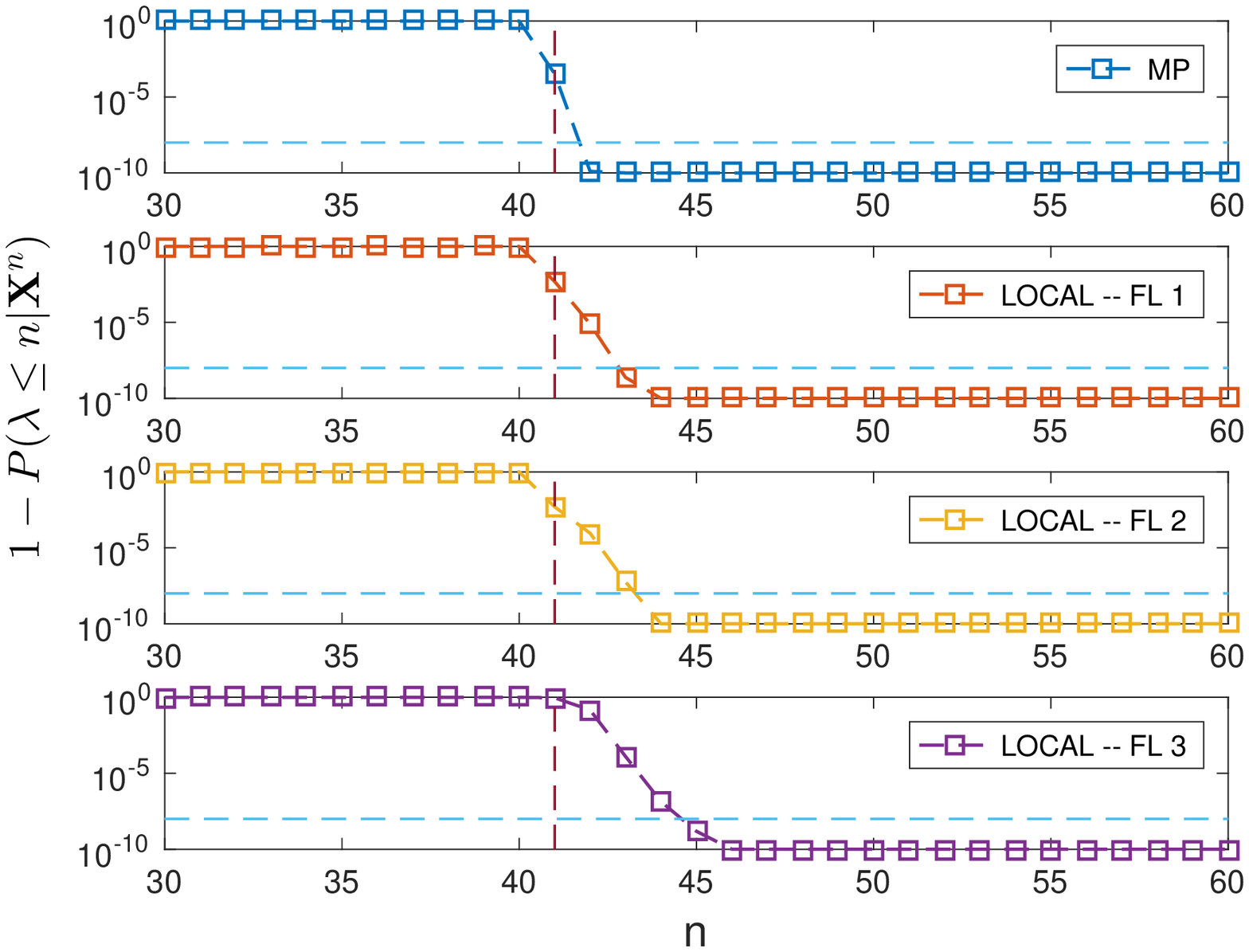}{The complementary cumulative distribution function $(1-P(\lambda\leq n|\mathbf{x}_*^n))$ of \textit{minimum} detector $\min\{\lambda_1\}$ for the major damage pattern. The dashed horizontal line is the threshold $\alpha_{fa} = 10^{-8}$. The dashed vertical line is the true damage time $\lambda_1=41$.}{0.9\linewidth}


The detectors' performances for the major damage are summarized in Fig.~\ref{fig:exp_probLike}. When the complementary cumulative distribution function (CCDF) crosses the threshold, a fault is detected. For the proposed damage method (MP algorithm), the CCDF starts to decrease immediately after a fault occurs in the network. With a threshold of $\alpha_{fa} = 10^{-8}$, the MP method only requires one time step to detect the damage. The asymptotically optimal delay is $(-\log(0.999)+4.44+6.06+6.27)^{-1}|\log(10^{-8})| \simeq 0.96$, which is close to the actual delay. If we only use the DSFs collected locally, the LOCAL method discussed in Section~\ref{sec:multiple} requires more time to detect damage. For example, the sensor on the first floor, which is close to the damage location, needs one more time step to declare a damage event. In this experiment, the chunk size is $400$ samples and the sampling frequency is $200$Hz. Therefore, each time step represents two seconds. Two seconds are usually negligible in many applications. But this short interval is critical for warning structure users for potential damage or collapse. Its asymptotically optimal delay by only using the first floor sensor is $(-\log(0.999)+6.27)^{-1}|\log(10^{-8})| \simeq 2.57$. If we use the DSFs collected from the 3rd floor sensor, the detection delay is four time steps, or equivalently eight seconds.

\putFig{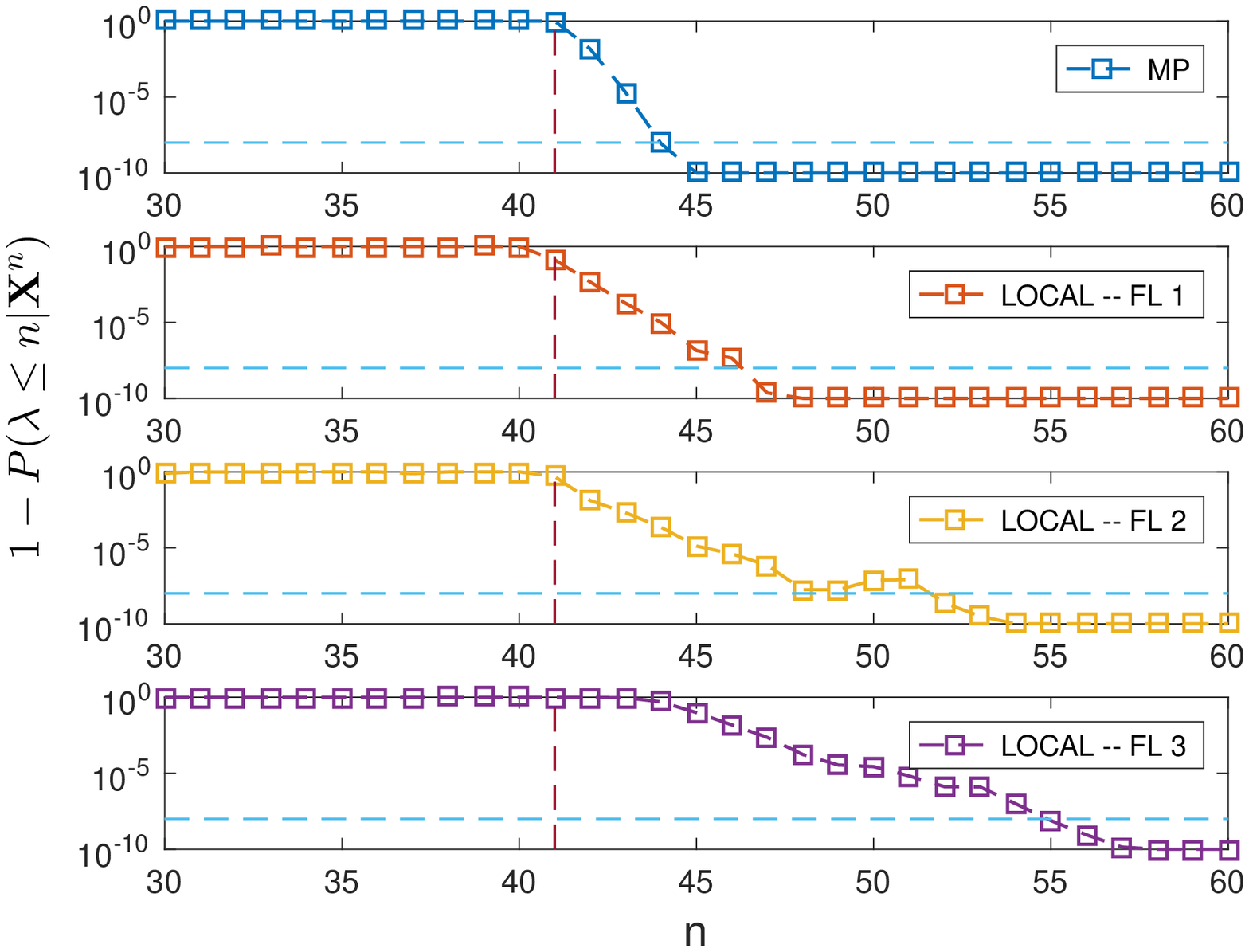}{The complementary cumulative distribution function $(1-P(\lambda\leq n|\mathbf{x}^n_*))$ of \textit{minimum} detector $\min\{\lambda_1\}$ for the minor damage pattern. The dashed horizontal line is the threshold $\alpha_{fa} = 10^{-8}$. The dashed vertical line is the true damage time $\lambda_1=41$.}{0.9\linewidth}

Fig.~\ref{fig:exp_probLike_minor} illustrates the CCDF for the minor damage (yielding) detection. The true damage time is still $41$. For MP, the damage delay is $3$ time steps, which is longer than the major damage detection. The main reason is that the KL distances in the minor damage are smaller. However, MP method still has shorter delay than the LOCAL method. This highlights that the MP method is robust to both major and minor damage.

\subsection{ASCE Benchmark Structure Simulation Validation}
\label{sec:sim_data}

\subsubsection{Description of Simulation}
\label{sec:des_sim}
In the previous section, only one damaged component is introduced to the structure. Our proposed algorithm can detect multiple damage simultaneously. To validate the performance of the proposed detectors on multiple damaged components, a large set of simulated data are obtained from the ASCE benchmark structure \cite{johnson2004phase}. The benchmark structure is a four-story two-bay by two-bay steel braced frame, as shown in Fig.~\ref{fig:ASCE}. The details about the benchmark structure and the simulator are provided in \cite{johnson2000benchmark}. The white noise excitation is applied on each floor along the y-axis. In order to get the responses associated with multiple simultaneous damage, we introduce the following damage patterns (DPs):
\begin{itemize}
\item DP0: no damage
\item DP1: all the braces of the 1st floor are removed
\item DP2: all the braces of the 3rd floor are removed
\item DP3: all the braces of the 1st and 3rd floors are removed.
\end{itemize}

In the ASCE benchmark structure, all the masses are loaded symmetrically. In addition, after removing the braces, the structure remains symmetric. Therefore, rather than using the data collected by all the sensors, in this study, we only use the responses collected by sensors $2, 6, 10$ and $14$. Because the damage patterns are introduced to each floor, we assign one damage variable to each floor, e.g. $\lambda_j$ for the $j$th floor. As the damage patterns only happen on the 1st and 3rd floors, we assume that only $\lambda_1$ and $\lambda_3$ will be the active damage variables in this test. 


\putFig{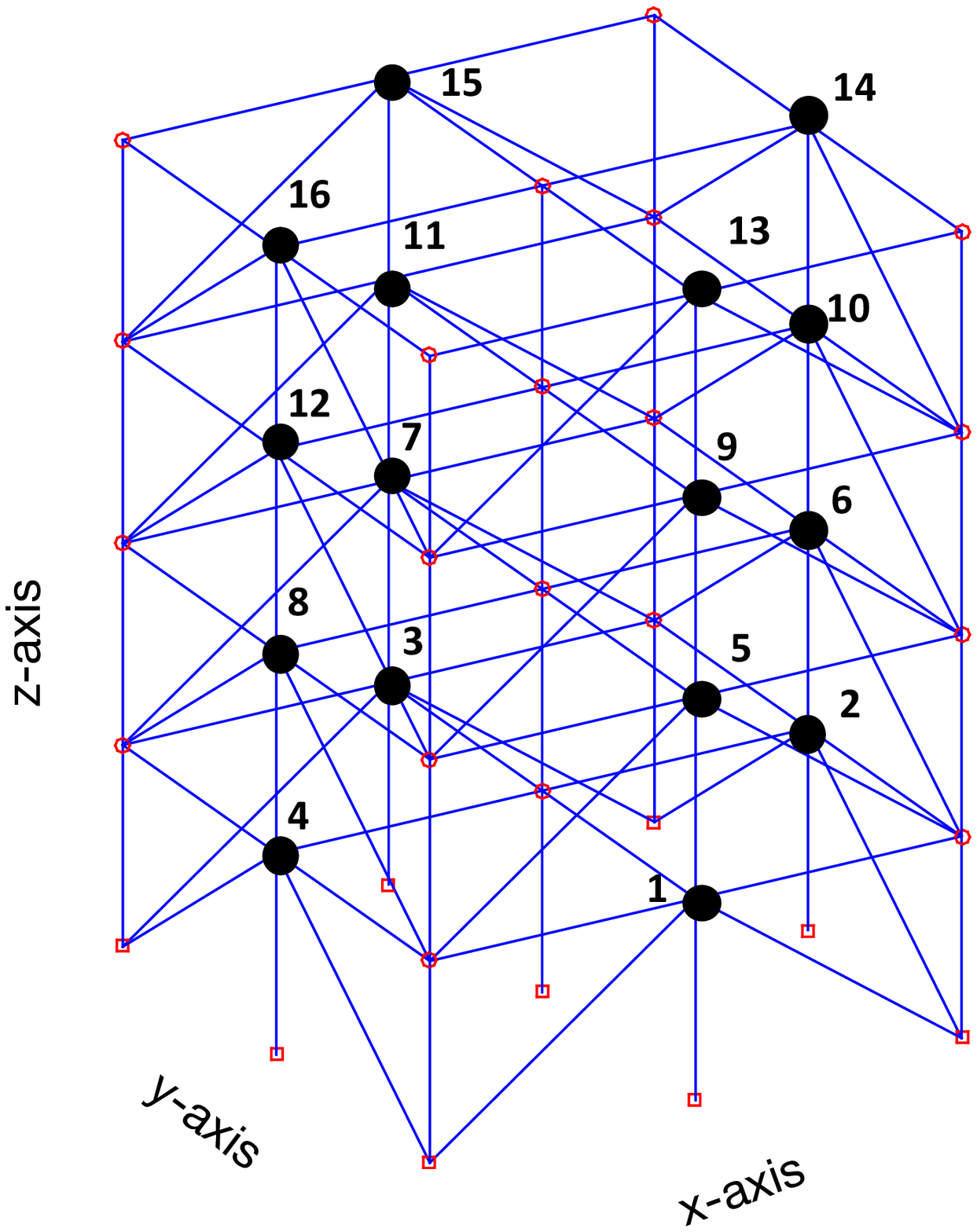}{Diagram of the ASCE benchmark structure with sensors (black dots).}{0.7\linewidth}

In the graphical model, a sensor is represented as a node. After assigning the damage variables ($\lambda$), we need to investigate the statistical dependencies (the edges in the graphical model) among the damage variables $\lambda$ and DSF variables ($\mathbf{X}$). Fig.~\ref{fig:AR_box} shows the box plots of the DSFs with different damage patterns. From this figure, we can observe that $\lambda_1$ has significant effects on sensor $2$ and sensor $6$. $\lambda_3$ affects the DSFs of all the sensors. Although $\lambda_2$ and $\lambda_4$ are not triggered by the defined DPs, we still assume it may affect the sensors installed one floor above and below. Based on these observations, we form the local domains and kernels in Table.~\ref{tab:ASCE_sensor}. The graphical model is shown in Fig.~\ref{fig:gm_asce}. 

\putFig{AR_box}{Box plot of the 1st AR coefficient $\theta_1$ for different damage patterns.}{0.9\linewidth}

\putFig{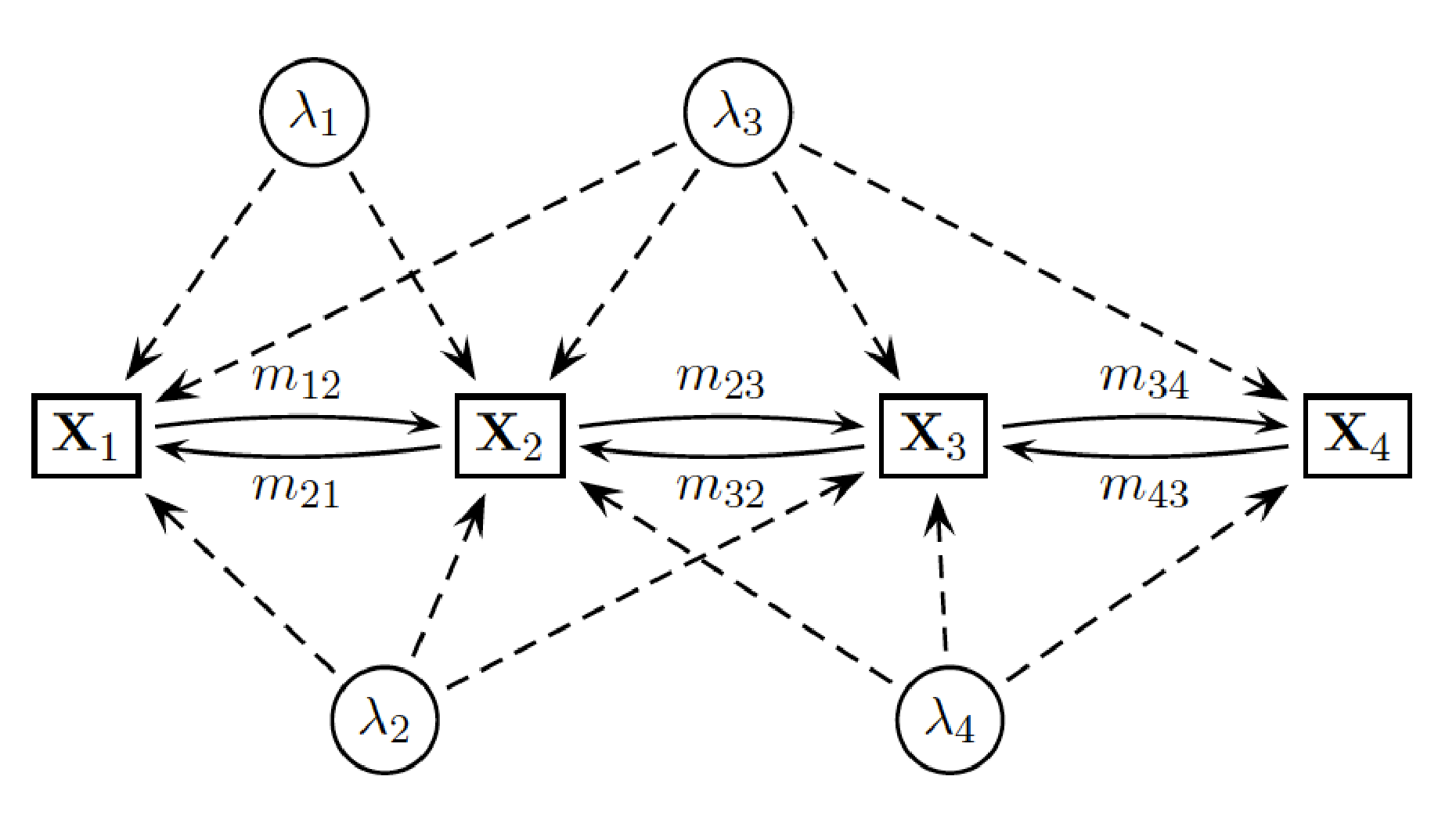}{Graphical model of damage variables ($\lambda$) and DSF variables ($\mathbf{X}$) in the ASCE benchmark structure. The dash arrows show the statistical dependencies and the solid arrows indicate the message passings.}{0.9\linewidth}

\begin{table}[h!]
\caption{Local domains and local kernels of the ASCE benchmark structure}
\label{tab:ASCE_sensor}
\centering
\begin{tabular}{c| l |l}
\hline
sensor & local domain & local kernel \\
\hline
2 & $\left\{\lambda_1,\lambda_2,\lambda_3\right\}$ & $\pi_1(\lambda_1)P(\mathbf{X}_2|\lambda_1,\lambda_2,\lambda_3)$ \\
\hline
6 & $\left\{\lambda_1,\lambda_2,\lambda_3,\lambda_4\right\}$ & $\pi_2(\lambda_2)P(\mathbf{X}_6|\lambda_1,\lambda_2,\lambda_3,\lambda_4)$ \\
\hline
10 & $\left\{\lambda_2,\lambda_3,\lambda_4\right\}$ & $\pi_3(\lambda_3)P(\mathbf{X}_{10}|\lambda_2,\lambda_3,\lambda_4)$  \\
\hline
14 & $\left\{\lambda_3,\lambda_4\right\}$ & $\pi_4(\lambda_4)P(\mathbf{X}_{14}|\lambda_3,\lambda_4)$ \\
\hline
\end{tabular}
\end{table}

\subsubsection{Results and Discussion}

\putFig{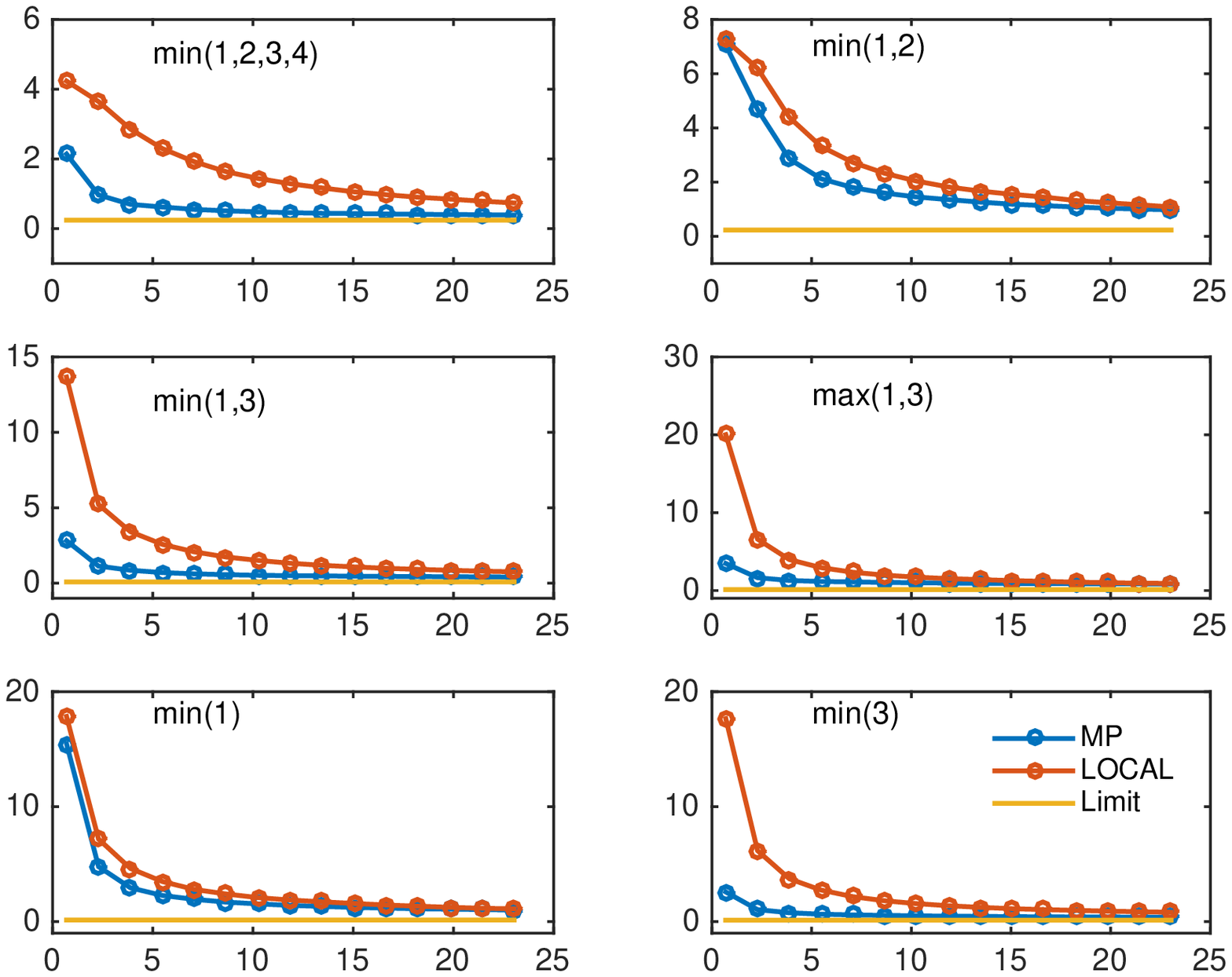}{Plots of the slope $\frac{1}{|\log\alpha_{fa}|}E\left[\tau_\mathbf{S}-\phi|\tau_\mathbf{S} \geq \phi\right]$ against $|\log\alpha_{fa}|$ for MP algorithm and LOCAL algorithm which only uses local data. False alarm rate $\alpha_{fa}$ ranges in $[0.5,10^{-10}]$.}{1.1\linewidth}

Fig.~\ref{fig:sim_MP_delay} shows the plots of the expected delay against the probability of false alarm $\alpha_{fa}$. The plots are generated by Monte Carlo simulation over $1000$ replications. All prior distributions are geometric distributions with $\rho_j = 0.05$. For the LOCAL algorithm, we apply detection rules on Sensor 6 since its local domain contains all four damage variables.

In Fig.~\ref{fig:sim_MP_delay}, we can observe that the MP algorithm outperforms the LOCAL algorithm. For detecting one fault (e.g., $\min(3)$), the MP algorithm, which utilizes all available data in the network, has a clear advantage over the LOCAL algorithm, for high to relatively low false alarm values around $\alpha = 10^{-5}$. Also, the advantage of MP over LOCAL is more emphasized when detecting multiple faults. It is also interesting to note that while MP converges to the asymptotically optimal bound, LOCAL seem to converge to a higher bound. Moreover, the \textit{minimum} rule has a shorter expected detection delay than the \textit{maximum} rule. As explained in Section.~\ref{sec:opt}, the \textit{minimum} rule has a lower theoretical limit.

\putFig{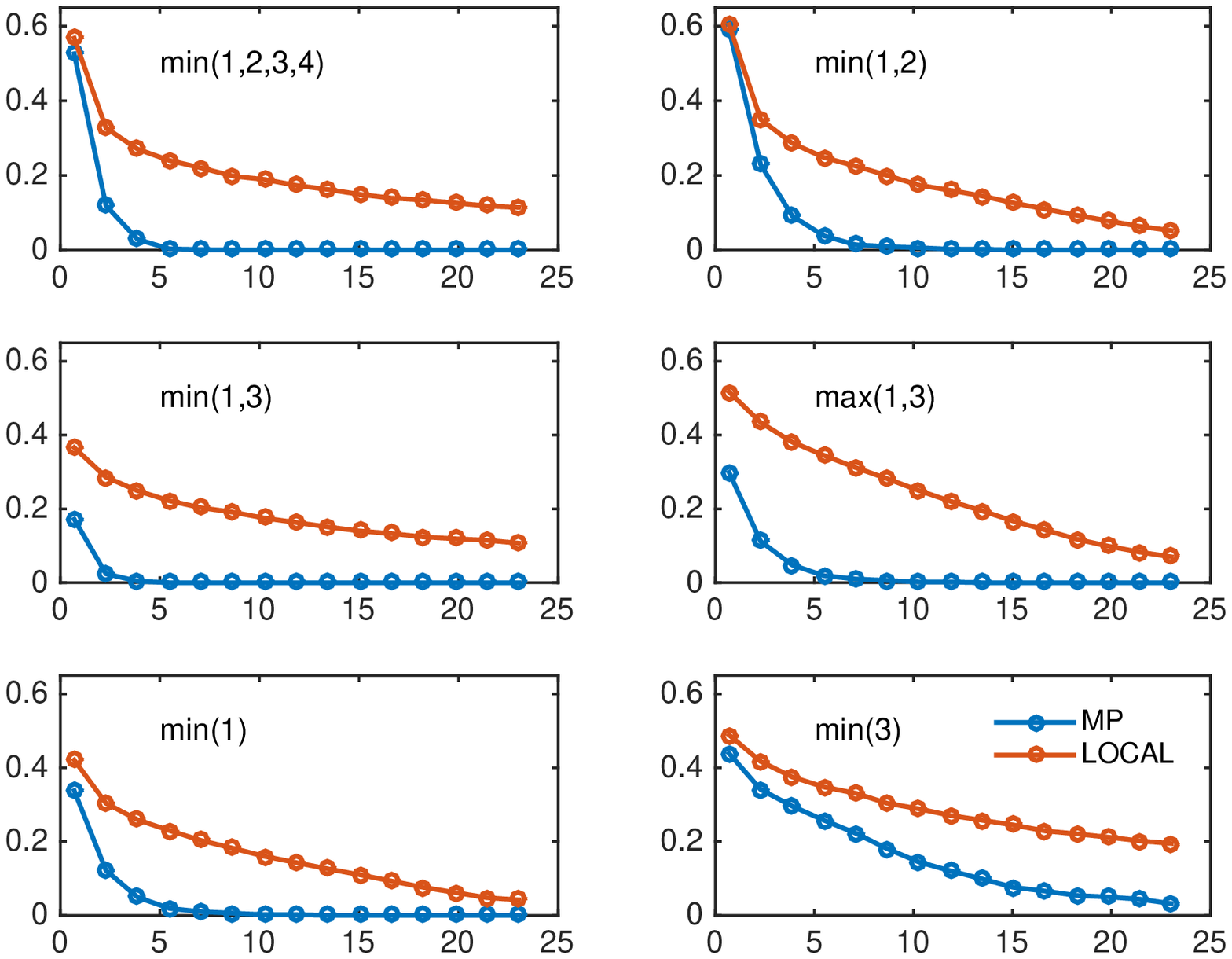}{Plots of the observed false alarm rate against $-\log\alpha_{fa}$ for MP algorithm and LOCAL algorithm which only uses local data. False alarm rate $\alpha_{fa}$ ranges in $[0.5,10^{-10}]$.}{1.1\linewidth}

Fig.~\ref{fig:sim_MP_fa} compares the empirical false alarm rate with the theoretical false alarm rate $\alpha_{fa}$ with the same setting. We observe that MP outperforms LOCAL in all cases. When $\alpha_{fa}$ is small, MP has no detection error approximately. However, the LOCAL algorithm has a detection error as large as $20\%$. This observation shows that utilizing all data in the network can significantly reduce the probability of false alarm.

\section{Conclusions}
\label{sec:conclusion}
In this paper, we present a data-driven algorithm to detect and localize multiple damage locations in a sequential and distributed manner. Specifically, we use random variables to represent the DSFs extracted from multiple sensors and the structural components we are interested in. Then, we use a probabilistic graphical model to connect these random variables based on their statistical dependence. After the graphical model formulation, we propose a group of detection rules that can be implemented sequentially and in a distribution fusion. Unlike many previous works, our method uses the DSFs from the entire WSN to diagnose structural health condition. Also, by utilizing the sparsity of graphical model, our algorithm does not require a base station and can detect damage at sensor devices. We use an experimental data set and a simulation data set to validate the consistency and robustness of this novel damage detection algorithm. Our results indicate that this method can detect and localize damage with high accuracy and low latency. Also, our analysis indicates that using all DSFs from the WSN can achieve a lower detection delay than only using local information.

\section*{Acknowledgement}
The authors would like to express their gratitude to Prof. Anne S. Kiremidjian from Stanford University for her comments and suggestions. The authors would also like to acknowledge Prof. Chin-Hsiung Loh from National Taiwan University, NCREE, and the ASCE Benchmark Committee for the data and MATLAB codes. This research was supported by NSF grants. Y. Liao is supported by the Charles H. Leavell Graduate Student Fellowship.

\bibliographystyle{IEEEtran}
\bibliography{citation}

\end{document}